\documentclass[twocolumn]{aastex62}

\usepackage{xcolor}
\usepackage{graphicx}
\usepackage{amssymb}
\usepackage{amsmath}

\newcommand\lal{Ly-$\alpha$ }
\newcommand\quasar{J0836+0054 }

\newcommand\dens{erg s$^{-1}$ cm$^{-2}$ \AA$^{-1}$}
\newcommand\fluxu{erg s$^{-1}$ cm$^{-2}$}

\newcommand{\HI}{{\rm H\,{\scriptstyle I}}}

\newcommand{\NV}{{\rm N\,{\scriptstyle V}}}
\newcommand{\SiII}{{\rm Si\,{\scriptstyle II}}}

\newcommand{\NHI}{N_{\rm {\scriptscriptstyle HI}}}

\graphicspath{{./}{figures/}}

\received{\today}

\submitjournal{ApJ}

\shorttitle{Three proximate LAEs at $z\sim5.8$}
\shortauthors{Bosman et al.}

\begin{document}

\title{Three Lyman-$\alpha$ emitting galaxies within a quasar proximity zone at $z\sim5.8$}

\correspondingauthor{Sarah E. I. Bosman}
\email{s.bosman@ucl.ac.uk}

\author{Sarah E. I. Bosman}
\affiliation{Department of Physics and Astronomy, University College London, London, UK}

\author{Koki Kakiichi}
\affiliation{Department of Physics and Astronomy, University College London, London, UK}
\affiliation{Department of Physics and Astronomy, University of California, Santa Barbara, USA}

\author{Romain A. Meyer}
\affiliation{Department of Physics and Astronomy, University College London, London, UK}

\author{Max Gronke}
\affiliation{Department of Physics and Astronomy, University of California, Santa Barbara, USA}

\author{Nicolas Laporte}
\affiliation{Kavli Institute for Cosmology, University of Cambridge, Cambridge, UK}
\affiliation{Cavendish Laboratory, University of Cambridge, Cambridge, UK}

\author{Richard S. Ellis}
\affiliation{Department of Physics and Astronomy, University College London, London, UK}

\begin{abstract}

Quasar proximity zones at $z>5.5$ correspond to over-dense and over-ionized environments. Galaxies found inside proximity zones can therefore display features which would otherwise be masked by absorption in the inter-galactic medium. We demonstrate the utility of this quasar-galaxy synergy by reporting the discovery of the first three `proximate Lyman-$\alpha$ emitters' (LAEs) within the proximity zone of quasar \quasar at $z=5.802$ (\textit{Aerith A, B} and \textit{C}).
\textit{Aerith A}, located behind the quasar with an impact parameter $D_\perp = 278$ pkpc, provides the first detection of a Lyman-$\alpha$ (Ly-$\alpha$) transverse proximity effect. We model the transmission and show it constrains the onset of J0836's quasar phase to $0.2 \text{Myr} < t < 20 \text{Myr}$ in the past.
The second object, \textit{Aerith B} at a distance $D=750$ pkpc from the quasar, displays a bright and broad  double-peaked \lal emission line. Based on relations calibrated at $z\leq3$, the peak separation implies a low ionizing $f_{\text{esc}} \lesssim 1\%$, the most direct such constraint on a reionization-era galaxy.
We fit the Ly-$\alpha$ line with an outflowing shell model, finding a completely typical central density $\text{log N}_{\text{HI}}/\text{cm}^{-2} = 19.3_{-0.2}^{+0.8}$, outflow velocity $v_{\text{out}} = 16_{-11}^{+4}$ km s$^{-1}$, and gas temperature $\text{log} T/\text{K} = 3.8_{-0.7}^{+0.8}$ compared to $2<z<3$ analogue LAEs.  
Finally, we detect an emission line at $\lambda=8177$ \AA\ in object \textit{Aerith C} which, if it is \lal at $z=5.726$, would correspond closely with the boundary of the quasar's proximity zone ($\Delta z < 0.02$ from the boundary) and suggests the quasar influences the IGM up to $\sim85$ cMpc away, making it the largest quasar proximity zone. 
Via the analyses conducted here, we illustrate how proximate LAEs offer unique insight into the ionizing properties of both quasars and galaxies during the epoch of reionization; we briefly discuss the prospects for finding further examples.
\end{abstract}


\section{Introduction} \label{sec:intro}

Reionization, the phase transition which rendered intergalactic hydrogen ionized, is thought to have concluded by $z\sim5.5$ \citep{Becker15, Greig17, Kulkarni19}. 
Mysteries regarding the morphology and driving sources of the process persist across a range of scales.

The unfolding of reionization during its end stages at $z\leq6.0$ has been tracked in great detail 
 using \lal transmission along the lines of sight to bright quasars \citep{Fan06, McGreer15, Eilers18, Bosman18}.  Correlations in inter-galactic medium (IGM) \lal opacity across $\gtrsim40$ proper Mpc (pMpc) have ruled out a homogeneous UV background (UVB) during the late stages of reionization, requiring the addition of mean free path and temperature fluctuations, an evolution of the global galaxy ionizing emissivity, and/or and increased contribution from rare ionizing sources \citep{Davies16, Chardin15,Keating16,Kulkarni19}. 

On small scales, faint galaxies with UV magnitude $M_{\text{UV}} > -18$ are expected to be the primary drivers of reionization \citep{Robertson13, Stark16, Dijkstra16}. 
The number of ionizing photons provided by a galaxy is the product of its ionizing emissivity, $\xi_{\text{ion}}$, and the escape fraction of these photons from the galaxy, $f_{\text{esc}}$. 
For galaxies with the same $M_{\text{UV}}$, models indicate that at least one of these parameters needs to be larger at $z>5.8$ than at $z\lesssim3$ in order for faint galaxies to provide the totality of the reionization photon budget \citep{Robertson15, Kakiichi18,Meyer19,Meyer20}.
However, the identification and study of early galaxies is complicated by the opacity of the IGM to wavelengths $\lambda<1215$\AA. Direct detection of Lyman-continuum (LyC) emission is currently only possible in the highly ionized IGM at $z\lesssim4$. 

Neutral hydrogen also hinders the use of the \lal emission line at $z>5$, the most common feature observable in the optical.
The number of continuum-selected Lyman-break galaxies (LBGs) which display \lal emission drops beyond $z>6$ \citep{Ouchi10}, most likely due to extended absorption wings in extremely neutral environments \citep{Dijkstra07, Laursen11,Mesinger15,Weinberger18}. Luminous galaxies are less affected by the decline in \lal visibility \citep{Santos16,Zheng17,Konno18, Mason18} owing to their probable location within early ionized bubbles which facilitates their observation \citep{Matthee15,Songaila18}. 
In addition, the observed shape of the \lal line is also affected by reionization.
At $z<4$, the \lal emission line occasionally displays a double-peaked shape whose morphology correlates with the presence of LyC leakage \citep{Verhamme15, Vanzella18, Izotov18} as well as a wide range of galactic properties \citep{Gronke17, Marchi18}. 
At $z>5$, the visibility of the blue peak of the \lal emission line is strongly suppressed \citep{Matthee17,Shibuya18}, limiting its usefulness. 

Currently, the only $z>5$ double-peaked \lal emitters (LAEs) are NEPLA4 at $z=6.55$ \citep{Songaila18, Mason18} and COLA1 at $z=6.59$ \citep{Hu16}.
Modelling of the \lal double peak in COLA1 and comparison with lower-$z$ analogues has yielded highly detailed information, including estimates of its inter-stellar medium (ISM) temperature $T\sim16000$K, a relatively low central neutral hydrogen density $N_{\text{HI}} \sim10^{17}$cm$^{-2}$, as well as dust opacity, velocity dispersion, and outflow speed \citep{Matthee18}. 
Further, the small velocity separations between the blue and red peaks of \lal in both COLA1 and NEPLA4 ($220$ km s$^{-1}$ and $300$ km s$^{-1}$ respectively) indicate an elevated $f_{\text{esc}} \gtrsim 0.1$ \citep{Izotov18}, suggesting these objects could be contributing to their own locally ionized environments. 
While such insight into reionization-era galaxies is invaluable, both of these galaxies are among the brightest at $z>6.5$ and display \lal luminosities $\sim7$ times higher than found in LAEs at $z\sim3$ and related analogues such as Green Pea (GP) galaxies (e.g.~\citealt{Yamada12,Yang17}). While they are tracing exceptionally highly-ionized regions during the epoch of reionization (EoR),
COLA1 and NEPLA4 may not be representative of the faint galaxies responsible for the bulk of the process.

Quasar proximity zones offer an alternative way of tracing over-ionized regions during reionization. Even in the significantly neutral IGM at $z>5.5$, luminous quasars are observed to be surrounded by ionized H~{\small{II}} regions sustained by the ionizing radiation from the active galactic nucleus (AGN; e.g.~\citealt{Madau00, Wyithe05, Bolton11}). 
In addition, UV-selected bright quasars ($M_{\text{AB}} < -27$) at $5.5<z<6.0$ reside in highly star-forming host galaxies \citep{Walter09} with large reservoirs of molecular gas (\citealt{Decarli18} and therein), and 
are expected to be hosted in dark matter haloes of masses $M_h \gtrsim10^{12}$ \citep{Shen07,Conroy13}. As such, the locations of EoR quasars should coincide with the most active sites of early galaxy formation (e.g.~\citealt{Overzier09}). 
This makes proximity zones ideal locations to observe reionization-era galaxies: \begin{enumerate} \item they correspond to over-dense environments; and \item the quasar clears the neutral hydrogen responsible for the attenuation of the \lal emission of galaxies, revealing features normally masked by IGM absorption. \end{enumerate}

In this paper, we demonstrate the power of galaxies found in quasar proximity zones, or \textit{proximate LAEs}, to constrain the ionizing properties of EoR galaxies and quasars.
We present the first three proximate LAEs: hereafter \textit{Aerith A, B} and \textit{C}. All three are observed to have unique properties related to their location. 

The observational data and derived physical properties of the objects are given in Section 2. In Section 3, we explore the morphology of the double-peaked LAE \textit{Aerith B} and model the \lal emission line using an expanding shell model. Section 4 discusses the structure of the proximity zone of J0836 inferred from a diagonally transverse proximity effect detected towards \textit{Aerith A}. As a proof of concept, modelling the transmission enables up to put limits on the timescale of the quasar's activity. We discuss implications for reionization in Section 5 and summarise in Section 6.

\begin{figure}[t]
\includegraphics[width=0.50\textwidth]{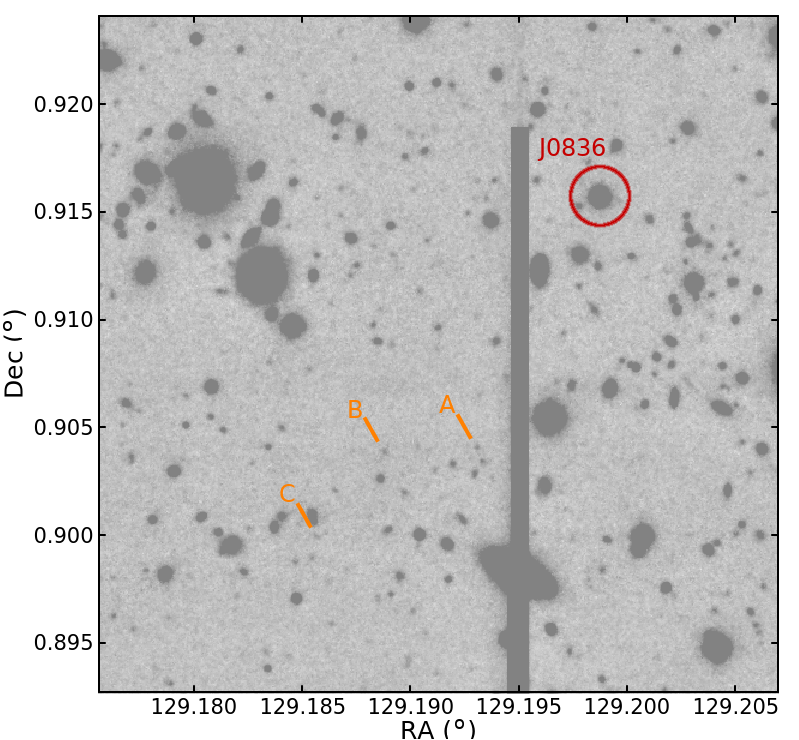}
\caption{Layout of the field around quasar J0836. The $z$ broad-band image is shown with contamination from a nearby star masked by a vertical rectangle.}
\label{fig:radec}
\end{figure}

Throughout the paper, we use a flat $\Lambda$CDM cosmology with $\Omega_m = 0.3089$ and $H_0 = 67.74$ \citep{Planck_latest}. Magnitudes are given in the AB system \citep{Oke83} and distances are quoted in proper distance units unless otherwise specified. We use $F$ and $\overline{F}$ to distinguish between measurements of flux and flux density (per \AA), respectively. Observational sensitivity uncertainties are denoted with {\small{(sys)}} while calibration uncertainties are indicated with {\small{(obs)}}. At $z=5.8$, $1' \simeq 0.24$ pMpc and a redshift interval $\Delta z = 0.05 \simeq 3.4$ pMpc.

\begin{figure*}[th]
\includegraphics[width=1.0\textwidth]{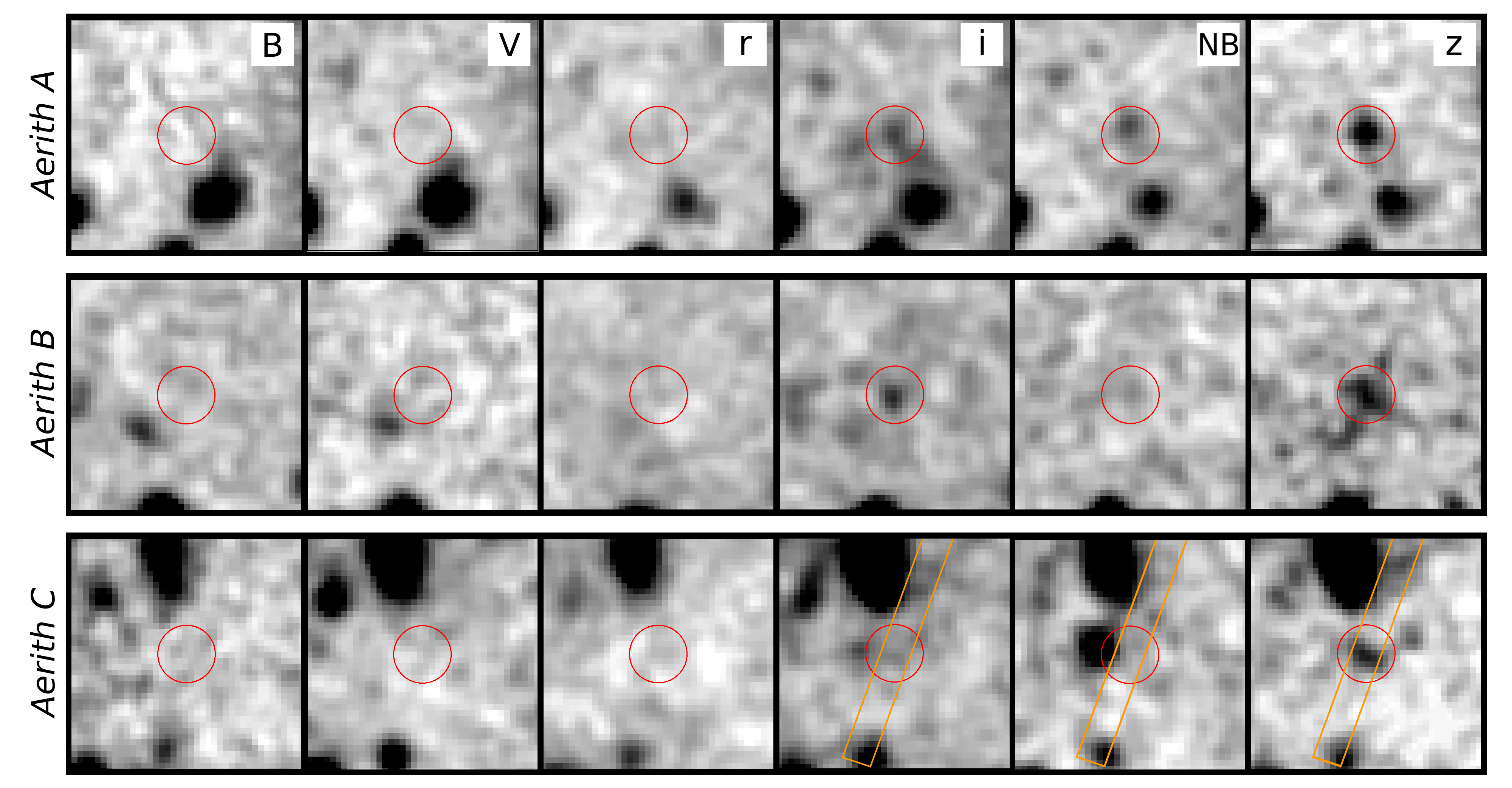}
\caption{Subaru/SuprimeCam photometry of the proximate LAEs. The narrow-band filter \textit{NB816} corresponds to $5.65 < z_{\text{Ly}\alpha} < 5.75$. A $10\sigma$ detection of \textit{Aerith A} can be seen in \textit{NB816}, which corresponds to emission bluewards of its $z_{\text{Ly}\alpha} = 5.856$. \textit{Aerith C} shows a physical offset between the peak emission in the $z$ band and the \textit{NB816} and $i$ bands of 6.3 pkpc assuming $z_{\text{Ly}\alpha} = 5.726$. The red circles are 3'' in radius and the DEIMOS slit position is shown in orange for \textit{Aerith C}.}
\label{fig:dropouts}
\end{figure*}

\begin{table}[t]
\begin{tabular}{l c c c c c c }
& $B$ & $V$ & $r$ & $i$ & $z$ & $NB816$ \\
\hline
$t_{\text{exp}}$ (s) & 6000 & 14400 & 3600 & 5920 & 9630 & 10800 \\
depth ($3\sigma$) & 28.2 & 28.4 & 27.9 & 27.6 & 26.4 & 27.2 \\
\end{tabular}
\caption{Photometry of the $34' \times 27'$ field around J0836 with Subaru-SuprimeCam. Magnitudes are calculated within a $2''$ aperture except for the $z$ broad-band which uses a $3''$ aperture.}
\label{table:exptime}
\end{table}

\begin{table*}
\center
\begin{tabular}{l l l l}
\hline
\hline
 & \textit{Aerith A} & \textit{Aerith B} & \textit{Aerith C} \\
 \hline
$\alpha_{\text{J2000}}$ & 08:36:45.24 & 08:36:46.28 & 08:36:47.04 \\
$\delta_{\text{J2000}}$ & 00:54:11.20 & 00:54:10.55 & 00:53:56.36 \\ 
$F_i$ ($10^{-17}$ \fluxu) & $7.8_{-1.7}^{+2.9}$ & $4.9_{-1.1}^{+1.9}$ & $3.9_{-0.8}^{+1.4}$ ${ }^{\text{(a)}}$ \\
$\overline{F_z}$ ($10^{-20}$ \dens) & $9.0\pm0.8 {\scriptstyle(\text{obs})} {}_{-1.3}^{+2.0} {\scriptstyle(\text{sys})}$ & $7.4\pm0.9 {\scriptstyle(\text{obs})} {}_{-1.0}^{+1.7} {\scriptstyle(\text{sys})}$ & $6.2\pm0.5 {\scriptstyle(\text{obs})} {}_{-0.9}^{+1.3}  {\scriptstyle(\text{sys})}$ \\
$F_{NB816}$ ($10^{-18}$ \fluxu) & $5.0\pm0.4 {\scriptstyle(\text{obs})} {}_{-1.1}^{+2.0}  {\scriptstyle(\text{sys})}$ & $<2.3$ & $23.0\pm1.0 {\scriptstyle(\text{obs})} {}_{-5.0}^{+9.0}  {\scriptstyle(\text{sys})}$ \\
$F_{\text{Ly}\alpha, \text{spec}}$ ($10^{-18}$ \fluxu) & $10.9\pm0.5 {\scriptstyle(\text{obs})} {}_{-0.6}^{+0.7}  {\scriptstyle(\text{sys})}$ & $30.1\pm0.5 {\scriptstyle(\text{obs})} {}_{-1.6}^{+1.8}  {\scriptstyle(\text{sys})}$ & $2.1\pm0.3 {\scriptstyle(\text{obs})} {}_{-0.2}^{+0.3}  {\scriptstyle(\text{sys})}$ \\
$z_{\text{Ly}\alpha}$ & $5.856 \pm 0.003$ & $5.793 \pm 0.003$ & $5.726 \pm 0.003$ \\
$D_\perp$ (pkpc) & 278 & 334 & 446 \\
$D_{\text{QSO}}$ (pkpc) & 3401 & 748 & 5219 \\
EW$_{\text{Ly}\alpha, \text{phot}}$ (\AA, rest) & $>10.1$ & $76_{-34}^{+55}$ & $55_{-5}^{+8}$ \\
EW$_{\text{Ly}\alpha, \text{spec}}$ (\AA, rest) & $18\pm4$ & $59_{-18}^{+28}$ & $5.0_{-0.5}^{+1.4}$ \\
\hline
$M_{\text{UV}}$ & $-21.3 \pm 0.2$ & $-21.0_{-0.2}^{+0.3} $ & $-20.8\pm0.2$ \\
$\text{log} L_{\text{Ly}\alpha}$ (erg s$^{-1}$) & $42.63_{-0.04}^{+0.05}$ & $43.03\pm0.03$ & $42.93_{-0.11}^{+0.15}$ ${ }^{\text{(a)}}$ \\
$\text{SFR}_{\text{UV}} (M_\odot\text{ yr}^{-1})$ & $24_{-4}^{+5}$ & $19_{-2}^{+5}$ & $16_{-2}^{+4}$ \\
$\text{SFR}_{\text{Ly}\alpha} (M_\odot\text{ yr}^{-1})$ & $49^{+22}_{-13}$ & $38^{+21}_{-14}$ & $32^{+18}_{-10}$ ${ }^{\text{(a)}}$ \\
\hline
\hline
\end{tabular}
\caption{Summary of important measurements and inferred quantities. Photometric rest-frame equivalent widths assume a flat spectral slope $\beta=-2$ and $z_{\text{Ly}\alpha}$. Inferred values for \textit{Aerith C} assume the $z$ band and \textit{NB816} detection are physically related. Limits are given at the $2\sigma$ level.\ ${ }^{\text{(a)}}$based on photometry. }
\label{table:megatable}
\end{table*}

\section{Observations} \label{sec:observations}

SDSS \quasar (J0836) is the third brightest quasar currently known at $z>5.7$ out of more than 305 objects\footnote{\url{http://www.sarahbosman.co.uk/list_of_all_quasars.htm}} \citep{Fan01, Banados16}. It is also the second most radio-loud quasar at $z>5.7$ out of 41 for which data is available \citep{Wang07, Banados15}. 
In this paper we make use of a 2.3h VLT/X-Shooter spectrum \citep{XSHOOTER} originally presented in \citet{McGreer15} and re-reduced in \citet{Bosman18}. 
Estimates of J0836's systemic redshift have varied widely in the literature (e.g.~\citealt{Stern03}: $z=5.774\pm0.003$; \citealt{Shen19}: $z=5.834\pm0.007$) due to the variety of methods used in the absence of detected molecular lines from the host galaxy \citep{Maiolino07}. Among optical and infrared emission lines, Mg~{\small{II}} 2800\AA \ most reliably traces the systemic redshift due to its empirical close agreement with host [C~{\small{II}}] 158 $\mu$m emission (e.g.~\citealt{Decarli18}). Unfortunately, Mg~{\small{II}} is heavily affected by atmospheric water absorption in our X-Shooter spectrum of J0836. Instead we use a combination of the O~{\small{II}} 1305\AA \ and C~{\small{II}} 1335 \AA \ emission lines as a proxy for Mg~{\small{II}}. Unlike other prominent optical/IR quasar emission lines used in redshift determination, O~{\small{II}} and C~{\small{II}} are not blends of multiple lines, and display no shift and negligible redshift scatter compared to Mg~{\small{II}} \citep{Meyer19-qso}. We determine the peak emission of the two lines using the QUICFit algorithm\footnote{https://github.com/rameyer/QUICFit}. This yields a redshift of $z_{\text{sys}} = 5.804 \pm 0.002$, which we will use throughout the paper.

In order to estimate the transmitted \lal flux inside J0836's proximity zone, we fit a physically-motivated emission model to its continuum and the Ly-$\alpha$, $\NV$ 1240\AA and $\SiII$ 1260\AA \ emission lines. A power-law is fitted to the continuum over wavelength intervals devoid of emission lines as in \citet{Bosman18}.  
We use a total of four Gaussian components to represent the broad and narrow components of \lal emission, and the single-component $\NV$\ and $\SiII$ lines. The emission line components are permitted to have a (single) velocity shift with respect to the quasar systemic redshift, which we find to be $\Delta v = 150\pm30$ km s$^{-1}$.

\subsection{SuprimeCam Photometry}

Observations with SuprimeCam on the 8.2m Subaru Telescope \citep{SUBARU, SUPRIMECAM} were conducted in a $34' \times 27'$ field of view around J0836 in 2004 (\citealt{Masaru06}; P.I.~Taniguchi). The field was imaged with the \textit{B, V, r+, i+, z+} (hereafter \textit{r,i,z}) broad-band filters as well as the narrow-band filter \textit{NB816} corresponding to \lal over $5.65<z_{\text{Ly}\alpha}<5.75$ ($50\%$ transmission bounds). The seeing in individual exposures was $<1.2''$ at all times. 
In another study, we aimed to identify LBGs around J0836 at $5.65 \leq z \leq5.90$ \citep{Meyer20}. 
We therefore initially selected candidates for spectroscopic follow-up based on their $r$, $i$ and $z$ magnitudes. 
Our selection criteria are described in \citet{Kakiichi18} and \citet{Meyer20} and briefly summarized here. 
We search an area of $\sim10'$ radius around the quasar for objects with $(r-i) > 1.0$ and $(i - z) < 1.0$ colors and a $3\sigma$ detection in the $z$ band. We then prioritise for follow-up the objects with $(r-i)>1.5$, no detections in the $r$ band, and finally use filler objects with narrow photometric redshift posteriors regardless of $(i-z)$ color. 
\citeauthor{Zheng06} (2006, hereafter Z06) used HST/ACS photometry and slightly different selection criteria to identify 7 $(i_{\text{ACS}}-z_{\text{ACS}})$ dropouts in the central $3'$ radius around J0836; we identify 19 candidates over the same area out of which 3 overlap with their selection (objects ``A'', ``B'' and ``F'' in their paper). Of these 19 candidates, we spectroscopically followed-up 11 including Z06's ``A'' and ``B'' (Section 2.2). 
Three of these (dubbed \textit{Aerith A, B} and \textit{C},) which lie in the proximity zone, form the basis of this paper. 
Their location with respect to J0836 is shown in Figure~\ref{fig:radec} and their spectral properties are introduced in Section 2.

Noting the importance of these 3 sources, we decided to carefully re-reduced the SuprimeCam archival observations to attain more accurate photometry in the $i$ and $z$ bands, and obtain measurements in the \textit{NB816} filter.
The re-reduction was carried out using the legacy pipeline {\tt SDFRED1} \citep{Yagi02, Ouchi04}. Magnitudes are extracted using {\tt SExtractor} \citep{Bertin96} and the limiting magnitude is estimated by distributing forced apertures in the $1'\times1'$ region surrounding the central quasar. We noticed infrared fringing around bright objects in the $z$ band which might have been affecting the photometry of \textit{Aerith A}, as well as a lower effective seeing in this band of $\sim 2''$. We therefore conservatively extract the total fluxes in $3''$ apertures, which do not contain any visible contaminating objects for our targets of interest (Figure~\ref{fig:dropouts}). Additionally, we mitigate the effect of fringing by masking the affected regions. This reduces the depth of the $z$-band photometry by $14\%$ at the location of \textit{Aerith A}. 
Zero-pointing of the photometry was carried out using 14 faint quasars and stars within the field of view with spectra available in the Sloan Digital Sky Survey Data Release 4 (SDSS DR4, \citealt{Blanton17}). We find this to be the dominant source of flux uncertainties, due to the non-linearity of the zero-point correction with magnitude and the relative lack of sufficiently faint standard sources. The measured scatter is of order $25\%$ at $2\sigma$ in all bands. Our measured sensitivities (Table~\ref{table:exptime}) are $\lesssim 1.5\sigma$ worse than those reported by \citeauthor{Masaru06} (2006, hereafter A06), who first presented the SuprimeCam observations, when accounting for the smaller $1''$ apertures used by those authors. Using a $1''$ aperture, we obtain flux measurements consistent at $1\sigma$ with A06 for \textit{Aerith A} and \textit{Aerith C} (A06's ``A'' and ``B'', respectively; Table \ref{table:megatable}) in all bands, except for $z(Aerith\ A)$ where we obtain $\overline{F_z} = 9.04 10^{-20}$ \dens, in closer agreement with Z06's HST/ACS values.

\subsection{DEIMOS spectroscopy}

Spectroscopic follow-up was conducted with the DEep Imaging Multi-Object Spectrograph (DEIMOS, \citealt{DEIMOS}) on the 10m Keck II telescope on March 7th and 8th, 2018, with the primary goal of confirming $5.65 \leq z \leq5.90$ LBGs candidates to use for cross-correlation with the \lal transmission towards J0836 \citep{Meyer20}. The total exposure time was 19000s (5.27h) with an average airmass of 1.099 (P.I.D.~U182, P.I.~Robertson). The DEIMOS field of view covers a slit mask area of $16.7'\times5'$ so that the central $5'\times5'$ area is entirely covered. Due to constraints in mask design, candidates were followed-up to maximize efficiency as well as based on their likelihood of being $5.65<z<5.90$ LBGs, also referred to as their grade. We targeted 32 targets in J0836's field.
We used a $1''$ slit with the 600ZD grating providing coverage over $4950 < \lambda(\text{\AA}) < 10000$ at a spectral resolution of 3.5\AA. Full details of our DEIMOS observations are given in \citet{Meyer20}.

The data was reduced with the DEIMOS \textit{DEEP2} Data Reduction Pipeline \citep{DEEP2-2, DEEP2-1} as well as with the open-source code \textit{Pypeit} \citep{pypeit} to check for consistency. 
In both cases, the reduction was performed in the standard way and taking slit losses into account. 
While extracted fluxes from the two reductions agree within $1\sigma$, the \textit{DEEP2} reductions achieve signal-to-noise ratios (SNR) 8\% larger on average and we use them in the rest of the paper. 
The search for lines was conducted visually by 5 of the authors (SEIB, RM, RSE, NL, KK) while being blind to individual targets' grades, photometric redshifts, and $y$-positions across the slit (which were scrambled to maximise mask efficiency). We identified 4 emission lines which were revealed to lie at the $y$-position of targeted dropouts. One target was consistent with an LAE at $z=5.284$ at a distance $d=39$ pMpc from the quasar line-of-sight, and is used in the analysis of \citet{Meyer20}. 

\subsection{Individual objects}

\subsubsection{Aerith A}
\begin{figure}[t]
\includegraphics[width=0.48\textwidth]{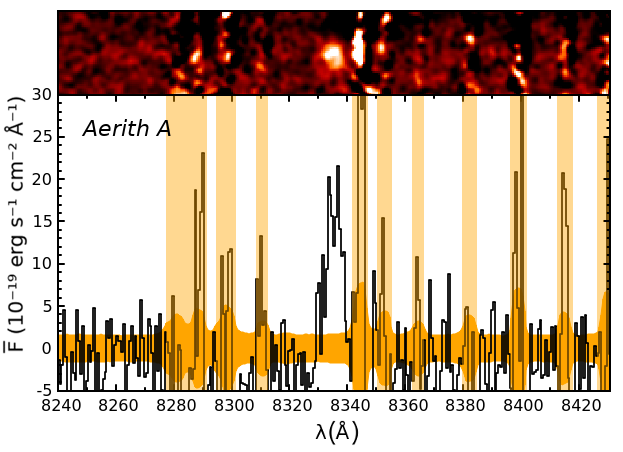}
\caption{DEIMOS spectroscopy of the emission line in \textit{Aerith A}. The lack of detection of other lines and a step in the detected continuum emission identify this as a \lal emission line at $z_{\text{Ly}\alpha} = 5.856$ (Section 2.3.1). In this Figure and all following, sky-lines are masked by vertical orange rectangles.}
\label{fig:lineA}
\end{figure}

At the location of {\textit{Aerith A}}, we detect an emission line at $\lambda = 8334.7$\AA \ with $\text{SNR}>20$ (Figure~\ref{fig:lineA}). No other emission lines are visible in the range $5000<\lambda<9550$\AA. A faint continuum is detected at $5.3\sigma$ redwards of the emission line, with intensity $\overline{F}_{\text{cont, spec}} = 8.5 \pm 1.6 {\scriptstyle(\text{obs})} \pm 0.6 {\scriptstyle(\text{sys})} \times 10^{-20}$ \dens over $9050<\lambda<9305$ \AA\. This is in agreement with the photometry in the $z$ band. 
No continuum is visible bluewards of the emission line. Taken together with the absence of other emission lines, this step in the continuum unambiguously identifies this object as a LBG and the emission line as \lal with $z_{\text{Ly}\alpha} = 5.856$. This corresponds to a distance $D_\parallel = 3.39$ pMpc behind the quasar.

The object is also detected at $10\sigma$ in the \textit{NB816} filter (Figure~\ref{fig:dropouts}, top panel). 
We stack the DEIMOS spectrum multiplied by the narrow-band transmission curve and obtain a corresponding measurement limit of $\overline{F}_{\text{NB, spec}} < 4.4 \times 10^{-20}$ \dens at $2\sigma$.

The $90\%$ transmission range of the \textit{NB816} filter is $8056<\lambda<8239$\AA, corresponding to shorter wavelengths than the \lal emission line of \textit{Aerith A}: this object  possesses a faint blue continuum. The window of transmission extends from $1.7$ pMpc to $11.8$ pMpc in front of the quasar such that $\geq34\%$ of the narrow-band width lies within the quasar's proximity zone. We conclude that this transmission is likely the result of significantly ionized foreground hydrogen: the quasar's proximity zone detected \textit{transversally} by a background LBG. Under this interpretation, we calculate a transmission integrated over the narrow-band $T = \overline{F}_{NB816} / k\overline{F}_z = 37\pm 6 {\scriptstyle(\text{sys})} {}_{-13}^{+22} {\scriptstyle(\text{sys})} \%$, where $k(\beta)$ is a continuum scaling factor depending on the spectral slope.\footnote{For $\beta=-2$ as we assume throughout, $k = 1.25$. For reference, $k=1.12$ for a slope $\beta=-1$.}
The non-detection of the continuum in the DEIMOS spectrum yields $T<39\%$ at $2\sigma$, for a combined constraint of $T = 30 \pm 11\%$. We therefore measure a \lal opacity over the narrow-band $\tau_{NB} = -\text{ln}(T)= 1.2_{-0.3}^{+0.4}$.

\subsubsection{Aerith B}

\begin{figure}[t]
\includegraphics[width=0.48\textwidth]{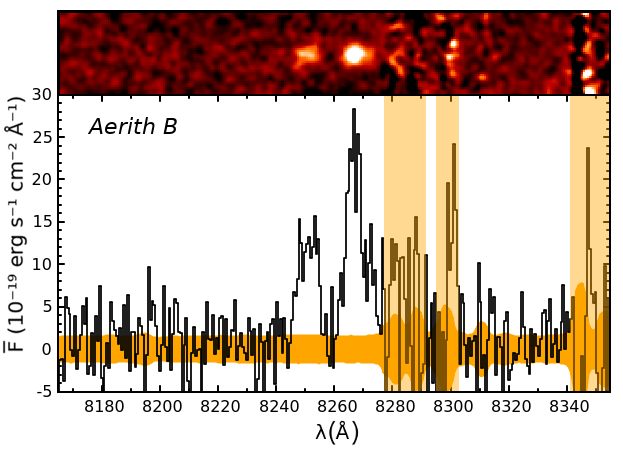}
\caption{DEIMOS spectroscopy of the two emission lines in \textit{Aerith B}. The velocity separation between the peaks ($580\pm80$ km s$^{-1}$) is incompatible with any common emission line doublets from a low-$z$ interloper. The double-peaked morphology is typical among double-peaked LAEs at $2<z<3$.}
\label{fig:lineB}
\end{figure}

\textit{Aerith B} displays two emission lines at $\lambda \lambda = 8251, 8267$\AA \ with a velocity separation of $\Delta v_{\text{sep}} = 580 \pm 80$ km s${}^{-1}$ (Figure~\ref{fig:lineB}). The separation is incompatible with an [O~{\small{II}}] $3727,3730$\AA \ doublet at $z=0.2282$ ($\Delta v_{\text{sep}} = 224$ km s${}^{-1}$), the only common emission line doublet with $\Delta v_{\text{sep}} < 1000$ km s${}^{-1}$. We measure the ratio of fluxes $A = F_{\text{blue}}/F_{\text{red}} = 0.65 \pm 0.05$. While some extreme [O~{\small{II}}] emitters do display similarly skewed emission ratios, none show such wide velocity separations \citep{Paulino-Afonso18}. We conclude that \textit{Aerith B} is a double-peak \lal emitter at $z_{\text{Ly}\alpha}\simeq5.793$. Using the decetion in the $z$ band to estimate the continuum, we measure the equivalent width of the two \lal emission components on either side of the trough at $\lambda = 8260$ \AA\ (Figure~\ref{fig:lineB}) as $W_{\text{red}} = 36_{-11}^{+17}$\AA; $W_{\text{blue}} = 23_{-7}^{+11}$ \AA.

Stacking the extracted spectrum over $9050<\lambda<9305$ \AA \ as for \textit{Aerith A} reveals a $4.6\sigma$ detection of $\overline{F}_{\text{cont, spec}} = 6.5 \pm 1.4 {\scriptstyle(\text{obs})} \pm 0.5 {\scriptstyle(\text{sys})} \times 10^{-20}$ \dens, in agreement with the $z$-band photometric detection. 
The object is not detected in \textit{NB816} at $2\sigma$ either photometrically or spectroscopically. 
We combine the two constraints into a loose upper limit $\overline{F}_{NB816} < 3.1 \times 10^{-20}$ \dens at $2\sigma$. The resulting fractional transmission over the narrow-band is $T = \overline{F}_{NB816} / k \overline{F}_z < 34\%$ at $2\sigma$, or $\tau_{NB} > 1.1$.

\subsubsection{Aerith C}

\begin{figure}[t]
\includegraphics[width=0.48\textwidth]{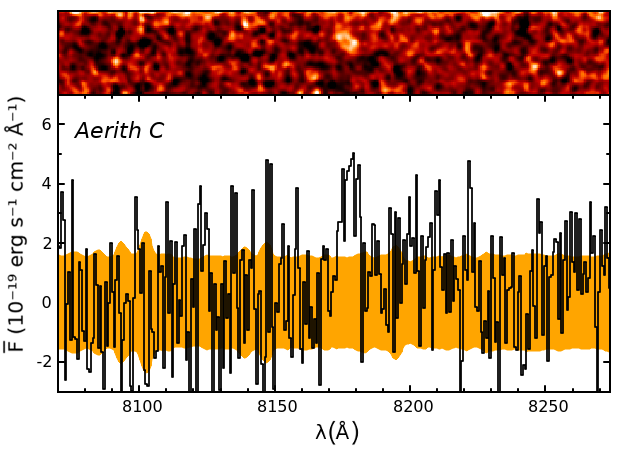}
\caption{DEIMOS spectroscopy of the emission line in \textit{Aerith C}. No other lines are detected in the spectrum. The physical offset  between the peak of $z$ continuum emission and \textit{NB816} emission (Figure~\ref{fig:radec}) suggests that the actual emission line flux is $\sim9$ times larger than captured in the DEIMOS slit (see Section 2.3.3).}
\label{fig:lineC}
\end{figure}

We detect an emission line at $\sim7\sigma$ above the noise at $\lambda  = 8176.9$\AA\ in the drop-out \textit{Aerith C} (Figure~\ref{fig:lineC}). Stacking the redward continuum yields a $2\sigma$ upper limit $\overline{F}_{\text{cont, spec}} <5.2$ \dens. This is in $2\sigma$ tension with the higher value from photometry. 

No other lines are detected in this spectrum, but the relative weakness of the $8176$\AA \ line makes ruling out an interloping object more challenging. Complementary evidence is provided by the \textit{NB816} image (Figure~\ref{fig:dropouts}, bottom panel). The \textit{NB816} wavelength range ideally encompasses the detected emission line. However, there is a large physical offset between the peaks of the $z$-band and \textit{NB816} emission, corresponding to $6.27$ pkpc at $z_{\text{Ly}\alpha}=5.726$. The positioning of the DEIMOS slit (orange rectangle, Figure~\ref{fig:dropouts}) was unfortunate in that the $z$-band continuum was centered while missing most of the \textit{NB816} emission. Indeed, the photometry indicates a much larger flux in the \textit{NB816} than we observed spectroscopically (Table~\ref{table:megatable}): only $\sim9\%$ of the emission line flux was recorded spectroscopically. This fraction is consistent with the detected emission line originating entirely from scattered light from the offset \textit{NB816} source.

No continuum is detected at the location of the offset \textit{NB816} detection in $B$, $V$, $r$ or $z$. 
A detection in the $i$ band is consistent with originating entirely from the emission line captured in \textit{NB816}. 
It is therefore likely that the totality of the \textit{NB816} flux originates in the emission line at $\lambda  = 8176.9$\AA. 
If this were a H$\alpha\  6465$\AA \ line at $z=0.245$, we would expect the corresponding H$\beta\  4862$\AA \ emission line to fall within the $r$ broad-band. No such detection is seen, with $F_r < 3.7 \times 10^{-18}$ \fluxu \ implying a line ratio $[\text{H} \alpha / \text{H} \beta] > 6.2$ at $2\sigma$. This is larger than seen in nearly all H$\alpha$ line emitters (e.g.~\citealt{concas19} and therein), thereby excluding the most common source of low-$z$ interlopers.

\begin{figure}[t!]
\includegraphics[width=0.45\textwidth]{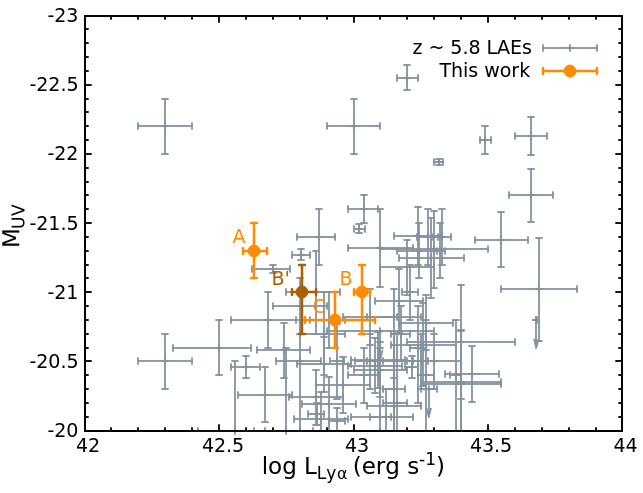}
\caption{Comparison of the \lal luminosities and $M_{\text{UV}}$ of the proximate LAEs with other bright ($M_{\text{UV}} < -20$) galaxies at $z>5.7$. \textit{Aerith A} displays a relatively low $L_{\text{Ly}\alpha}$ for its $M_{\text{UV}}$, which could indicate a spectral slope harder than $\beta=-2$. The darker symbol labeled \textit{B'} indicates the properties of \textit{Aerith B} if its blue \lal peak had been absorbed by the IGM. The comparison sample is drawn from \citet{Shibuya18, Mallery12, Hu10, Ouchi08, Matthee17, Ding17, Jiang13, Higuchi19, Jiang18}, with some values presented in \citet{Harikane19}.}
\label{fig:maglum}
\end{figure}

\citet{Masaru06} identified the \textit{NB816}-only source as a separate, related component to the $z$ detection. If the emission line at $\lambda  = 8176.9$ is \lal at $z=5.7263$, the $6.3$ pkpc physical offset between the $z$ continuum and the emission line is in excess of any objects previously reported, even in the cases of very clumpy high-$z$ galaxies \citep{Carniani18} when the UV and dust continuum are frequently offset from each other \citep{Maiolino15, Carniani17}. Various mechanisms including inhomogenous ionization in a galaxy are invoked at $z>6.0$ to explain the frequent offsets between continuum emission lines such as [C~{\small{II}}] 158$\mu\text{m}$ and highly ionized nebular lines such as [O~{\small{III}}] 5007 \AA \ \citep{Katz19}, but these offsets are $\leq3$ pkpc. Since the redshifts of the two components are consistent, another possibility that of two associated galaxies (potentially a galaxy merger) in which one member displays a very large Ly$\alpha$/UV ratio and the other a very small ratio. We speculate this could arise through an inhomogeneous/clumpy distribution of star formation and dust, as is sometimes seen in young galaxies (e.g.~\citealt{Carniani18} and therein).

\subsection{UV magnitudes and star formation rates}

\begin{figure}[t!]
\includegraphics[width=0.48\textwidth]{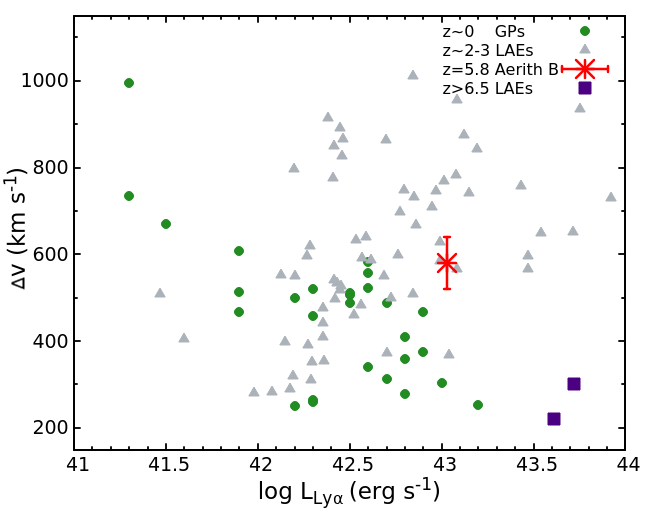}
\caption{The red-blue peak velocity separation in \textit{Aerith B} is consistent with the median seen in both $z\sim 2-3$ LAEs
and $z\sim0$ Green Pea galaxies \citep{Yang17}, although the latter tend to be $\sim0.5$ dex fainter in $L_{\text{Ly}\alpha}$ at the same peak separation. The $z>6$ double-peaked LAEs COLA1 and NEPLA4 are outliers compared to both comparison samples. 
The $z \sim 2-3$ sample is drawn from \citep{Yamada12, Kulas12, Hashimoto15, Vanzella16}. $L_{\text{Ly}\alpha}$ values from \citet{Kulas12} have been adjusted to reflect a Salpeter IMF rather than Chabrier.}
\label{fig:analogues}
\end{figure}

We calculate the UV magnitude $M_{\text{UV}}$ assuming a flat spectral slope $\beta = -2$ and the $k$-correction $-2.5(\beta+1) \text{log}_{10} (1+z_{\text{Ly}\alpha})$. $M_{\text{UV}}$ can be related to the star-formation rate (SFR) via $\text{SFR (}M_\odot\text{ yr}^{-1}\text{)} = 1.4 \times 10^{-28} L_{\nu, UV}$ \citep{Kennicutt98} assuming a Salpeter IMF \citep{Salpeter}. Alternatively, the SFR can be estimated within $15\%$ from the properties \lal emission line alone \citep{Sobral18, Sobral19}:
\begin{equation}
\text{SFR}_{\text{Ly}\alpha} \left[M_\odot\text{ yr}^{-1}\right] = \frac{L_{\text{Ly}\alpha} \times 7.9\times 10^{-42}}{(1 - \text{f}_{\text{esc}}) (0.042 \times \text{EW}_0)},
\end{equation}
where $\text{EW}_0$ is the equivalent width of the \lal line in the rest frame and a Salpeter IMF is again assumed. The resulting estimates of SFR are shown in Table~\ref{table:megatable} assuming $\text{f}_{\text{esc}} = 0.1$ for \textit{Aerith A} and \textit{Aerith C}, as measured in LAEs at later epochs \citep{Verhamme17,Fletcher18}. We use $\text{f}_{\text{esc}} = 0.05$ for \textit{Aerith B} due to the results of the \lal line fitting presented in Section 3. The $\sim2\sigma$ disagreement between SFR${}_{\text{UV}}$ and SFR${}_{\text{Ly}\alpha}$, with SFR${}_{\text{Ly}\alpha} \simeq 2 $ SFR${}_{\text{UV}}$, is common in $z\sim5.7$ UV-selected LBGs with $\text{SFR}\lesssim40 M_\odot \text{yr}^{-1}$ (e.g.~\citealt{Sobral19} and therein). Additional uncertainty in the UV SFR could be due to a spectral slope harder than $\beta=-2$. Alternatively, the \lal-derived SFR is sensitively dependent on the assumed shape of the initial mass function (IMF); a Chabrier IMF \citep{Chabrier} results in predictions a factor $\sim2$ lower compared to the Salpeter IMF assumed in equation (1).

Both the $M_{\text{UV}}$ and $L_{\text{Ly}\alpha}$ properties of our objects are typical of $z\sim5.8$ galaxies for which both measurements are available (Figure~\ref{fig:maglum}). Interestingly, \textit{Aerith C} displays a typical $L_{\text{Ly}\alpha} / M_{\text{UV}}$ ratio under the assumption that its UV and emission-line components are related despite the 6.3 pkpc physical offset.
\textit{Aerith A} displays a $L_{\text{Ly}\alpha}$ about 0.5 dex lower than the median value at $z\sim5.8$ given its UV magnitude. Its $L_{\text{Ly}\alpha} / M_{\text{UV}}$ ratio is comparable to the objects of \citet{Jiang13}, which possess particularly steep UV continuum slopes ($\beta\leq-2$). This could be indicative of a young stellar population and/or lack of dust. We also note that most samples of LAEs at $z\sim5.8$ are Ly$\alpha$-selected rather than UV-selected, which creates a sampling bias to higher values of $L_{\text{Ly}\alpha}$.


\begin{figure*}[t!]
\centering
\includegraphics[width=\textwidth]{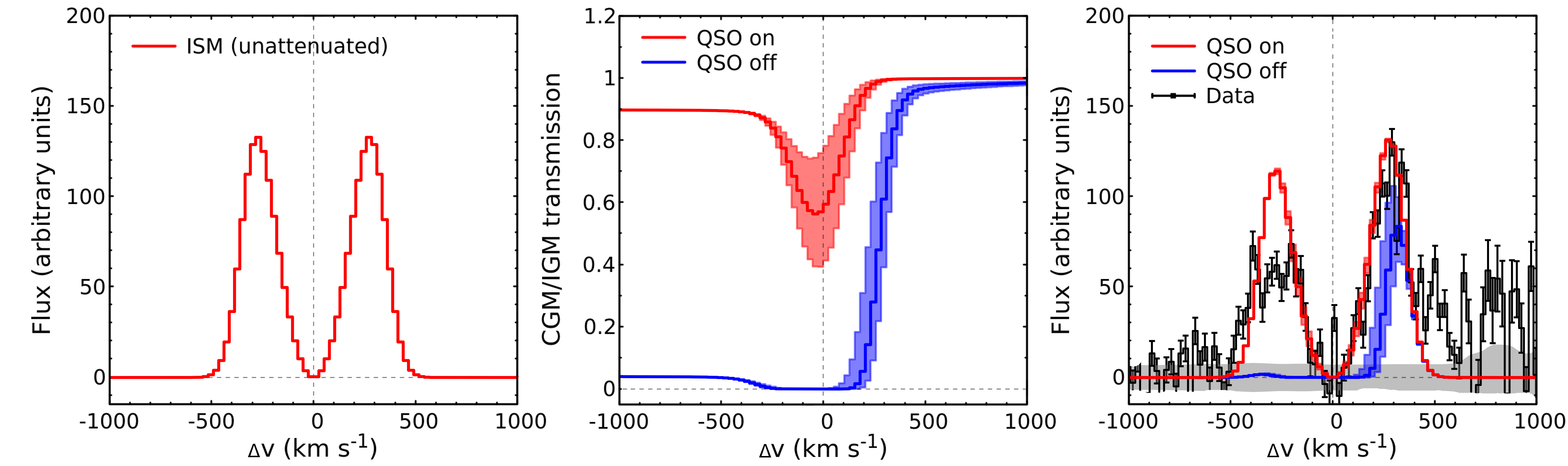}
\caption{CGM attenuation of the \lal double peak in \textit{Aerith B}. \textbf{Left}: Intrinsic \lal emission in the galaxy's ISM. \textbf{Middle}: Attenuation curves including the effect of \textit{Aerith B}'s CGM and IGM transmission resulting from J0836 and its clustered galaxy population (red) and the expected IGM attenuation in the absence of a luminous quasar (blue). \textbf{Right}: Resulting observed \lal emission line structure, with observations shown in black and models colors the same as the middle panel.}
\label{fig:spec-B}
\end{figure*}

\section{Aerith B: A Double-Peaked LAE and Constraints on the Escape Fraction at $z\sim6$}

The most striking feature of the newly-discovered proximate LAEs is the wide double-peaked \lal emission line in \textit{Aerith B}. Absorption by the partially neutral IGM makes this feature exceedingly rare at $z>5$ \citep{Hu16, Songaila18}, and destroys the large amount of information on galactic properties it contains. However, \textit{Aerith B} is located a distance $D_{\text{QSO}} = 748$ pkpc away from a $M_{\text{UV}} = -27.75$ quasar \citep{Banados16}. The quasar's contribution to the local ionization field ($\Gamma_\text{HI}$) is $\gtrsim 10$ times larger than the radiation field due to the galaxy itself, and $\sim 5$ times stronger than the UVB at its peak at $z=2$ \citep{Bolton07, Becker07, Faucher08-UVB}. 

We show, in this Section, how this proximate LAE offers unique insight into the escape fraction of ionising photons at $z\sim6$.

\subsection{Visibility of the \lal emission line at $z>5.5$}

The \textit{Aerith B} galaxy provide direct evidence that the $z>5.5$ IGM is affecting the visibility of the Ly$\alpha$ line. This effect has been commonly argued to be responsible for the declining Ly$\alpha$ fraction in LBGs at $z>6$ \citep{Stark10,Pentericci14, Mason18}, the sharp decline of the LAE number density \citep{Choudhury15, Weinberger18} and changes in the clustering of LAEs \citep{Furlanetto06, McQuinn07, Ouchi18}.

The velocity separation between the blue and red peaks of \lal in \textit{Aerith B} is $\Delta v_{\text{sep}} = 580 \pm 80$ km s${}^{-1}$, comparable to the median for double-peaked $z\sim3$ LAEs and for double-peaked $z\sim0.3$ GPs (Figure~\ref{fig:analogues}). Interestingly, current surveys of LAEs at $z\sim5.7$ are nearly all sensitive at this level, both in terms of SNR and spectral resolution. 
For example, 44 LAEs within a potential proto-cluster at $z\sim5.7$ compiled by \citet{Harikane19} were all observed with spectroscopic resolutions no worse than $\Delta v = 300$ km s${}^{-1}$, and have fainter $L_{\text{Ly}\alpha}$ than \textit{Aerith B}, but none were found to display a double-peaked \lal line. Confusion with the [O~{\small{II}}]$3727$\AA \ doublet ($\Delta v = 224$ km s${}^{-1}$) is 
potentially an issue for identifying the double \lal peak when combined with the lack of available optical lines to rule out interlopers at $z>4.8$, but this is 
negligible in the regime of $\Delta v>500$ km s${}^{-1}$. At $2<z<3.2$, roughly $15-25\%$ of all LAEs display double-peaked \lal lines at least as widely separated as this \citep{Kulas12, Trainor15}. Similarly, $\sim25\%$ of $z=0$ analogues posses such wide peak separations \citep{Rivera-Thorsen15, Yang17}.
The lack of $\Delta v > 500$ km s${}^{-1}$ double-peaked LAEs detected at $z>5.5$ compared to $z<3$ is therefore likely due to reasons other than observational completeness, such as absorption by the IGM. It is still surprising that two double-peaked LAEs should be found outside of proximity zones at $z>6.5$, but none at $5<z<6.5$, where the completeness is much higher.

The visibility of the \lal blue peak in the $z>6.5$ LAEs NEPLA4 and COLA1 is speculated to arise from local `ionized bubbles' sourced by the galaxies themselves and/or an associated highly-leaking population (e.g.~\citealt{Matthee18}). However, both of those galaxies are significantly brighter than \textit{Aerith B} and the galaxies in \citet{Harikane19} (Figure~\ref{fig:analogues}). Since \textit{Aerith B} demonstrates that $z>5.5$ LAEs do sometimes possess intrinsically double-peaked \lal profiles, 
we conclude that moderately bright galaxies are not generally able to sustain their own significantly ionized bubbles even in the context of a $z\sim5.7$ proto-cluster \citep{Harikane19}.

\subsection{Ionizing escape fraction}

The \lal peak separation is a highly sensitive tracer of the ionizing escape fraction, well-calibrated on studies of $0\leq z\leq 3$ galaxies with detections of LyC emission \citep{Jaskot13, Hayes15, Izotov18}. Unlike COLA1 and NEPLA4, \textit{Aerith B}'s \lal peak separation is wide and typical for $0<z<3$ LAEs with the same \lal luminosity (Figure~\ref{fig:analogues}).
The wide Ly$\alpha$ peak separation favours a low $f_{\rm esc}(\rm LyC)$. If we adopt the empirical fitting formula of \citet{Izotov18}, 
\begin{equation}
f_{\rm esc}({\rm LyC})=3.23\times10^4 \Delta v_{\rm sep}^{-2}-1.05\times10^2 \Delta v_{\rm sep}^{-1}+0.095,
\end{equation} then for the measured peak separation of Aerith B $\Delta v_{\rm sep}=580\pm80\rm~km~s^{-1}$ we find a LyC escape fraction,
\begin{equation}
f_{\rm esc}({\rm LyC})\approx0.01~~(\mbox{for Aerith B $z\simeq5.79$}).
\end{equation}
Since direct LyC detection in objects with such a wide Ly$\alpha$ peak separation is rare in low-$z$ analogues and a double peak typically indicates a high $\HI$ column density of the ISM, this value should be considered an upper estimate. 
Theory predicts the preferential escape frequency of \lal photons, and thus, the peak separation to be governed by the H~{\small{I}} column density \citep{Adams72, Neufeld90}. 
Simulations indicates that a such wide peak separation is suggestive of the absence of low ($\NHI<10^{17}\rm~cm^{-2}$) column density channels in the system \citep{Kimm19, Kakiichi19}.

In contrast, the double-peaked \lal emission of COLA1 at $z=6.59$ has a separation $\Delta v_{\rm sep}=220\pm20\rm~km~s^{-1}$ indicating \citep{Matthee18}:
\begin{equation}
f_{\rm esc}({\rm LyC})\approx0.29~~(\mbox{for COLA1 $z\simeq6.59$}).
\end{equation}
The narrow peak separation in COLA1 indicates the presence of low column density channels through which ionizing radiation can freely escape, but high enough ($>10^{14}\rm~cm^{-2}$) that scattering of Ly$\alpha$ photons can still take place at the core of the system. 
The difference between the Ly$\alpha$ lines in \textit{Aerith B} and COLA1 and NEPLA4 highlights that a variety of escape fractions are present at the tail end of reionization. The low $f_{\text{esc}}$ in \textit{Aerith B} is unlikely to be due to viewing angle, as simulations predict that \lal peak separations $\Delta v>300$ km s$^{-1}$ require an significant absence of ionizing channels \citep{Kakiichi19}. Rather, the difference could be due to intrinsic luminosity and/or clustering, with COLA1 and NEPLA4 being 0.5 dex brighter than \textit{Aerith B} and potentially residing in ionized bubbles they contribute to sustaining.

The flux ratio between the two peaks of \lal in \textit{Aerith B}, $A = F_{\text{blue}}/F_{\text{red}} = 0.65 \pm 0.05$, is in good agreement with the $A / W_{\text{Ly}\alpha}$ relation suggested by \citet{Erb14}. Those authors showed that in LAEs at $2<z<3$, $W_{\text{Ly}\alpha}$ anti-correlates with a systematic velocity offset between the peak of \lal emission and nebular lines. 
For $W_{\text{Ly}\alpha} \sim 60$\AA, a typical (extreme) offset is $150$ ($300$) km s${}^{-1}$. 
We are able to accurately obtain the systemic redshift of \textit{Aerith B} by using the minimum between the two \lal peaks, which traces the gas responsible for \lal scattering. Indeed, the offset between this point and the peak of \lal emission is $\sim250$ km s${}^{-1}$, in agreement with lower-$z$ results. 
The peak height ratio being different from 1 is indicative either of attenuation by the CGM and IGM, or an outflowing shell of material \citep{Bonilha79}, or more likely both (Figure~\ref{fig:spec-B}). In order to extract further physical information on the galaxy, we must disentangle these effects.

\subsection{CGM and IGM attenuation of the line}

How much of the peak asymmetry in \textit{Aerith B} could be due to CGM and IGM attenuation? Cross-correlation measurements between LAEs and the \lal forest in multiple quasar fields \citep{Meyer19, Meyer20} show evidence for CGM attenuation of \lal transmission around LAEs (or related metal tracers) on $\lesssim1\rm~pMpc$ scales at $z\sim6$ which should be taken into account to determine the intrinsic properties of the galaxy. 

We model the effect of CGM attenuation as follows. The transverse \lal absorption by the CGM links the mean line-of-sight effective optical depth $\tau_{\rm{eff}}^{\rm Ly\alpha}$ to the \lal emission line of a galaxy schematically via \citep{Kakiichi18b},
\begin{align}
&\tau_{\rm{eff}}^{\rm Ly\alpha}(\nu_e, r_\perp)\propto \\
&\int_{-\infty}^{\infty} dv\int_0^\infty d\NHI f(\NHI) \left[1+\xi_v(v,r_\perp)\right]\left[1-e^{-\sigma_\nu\NHI}\right],\nonumber
\end{align} 
with $\sigma_\nu=\sigma_\alpha\phi_V\left[\nu_e\left(1-v/c\right)\right]$ where $\phi_V$ is the Voigt profile, $f(\NHI)$ is the $\HI$ column density distribution function, and $\xi_v(v,r_\perp)$ is the velocity-space correlation function between galaxies and \lal absorbers. The interested reader can refer to the original paper for details. Two parameters choices are important, as the CGM absorption depends on the average velocity scatter of the absorbing gas ($\sigma_\alpha$) and the innermost radius of absorption $r_{\text{min}}$. We arbitrarily fix $\sigma_\alpha$ to an typical value for low-$z$ LAEs of $100$km s$^{-1}$ \citep{Gronke17}. The mass of \textit{Aerith B} is estimated around $\sim10^{11} M_\odot$ (see discussion in Section 5.1), which corresponds to a virial radius of $\sim20$ pkpc at $z=5.8$. We therefore produce a range of curves for $r_{\text{min}} = 10, 20$ and $50$ pkpc, resulting in the uncertainty in CGM absorption shown in the middle panel of Figure~\ref{fig:spec-B}.

We set the normalization of $f(\NHI)$ by requiring $e^{-\tau_{\rm{eff}}^{\rm Ly\alpha}(\nu_e)}$ to asymptotically approach $\langle e^{-\tau_\alpha(r_{\perp},r_\parallel)}\rangle$ bluewards of the line center in order to recover the correct limit of IGM attenuation. The resultant $e^{-\tau_{\rm{eff}}^{\rm Ly\alpha}(\nu_e)}$ then gives the mean estimate of the CGM+IGM attenuation curve around the \lal emission line of a galaxy.

The CGM+IGM attenuation curve near the \lal line profile at the position of {\it Aerith B}, as indicated in Figure~\ref{fig:spec-B}, shows that the strong ionizing radiation field ($\Gamma_{\rm HI}\sim10^{10}\rm~s^{-1}$) from the quasar is needed to raise the blue transmission of \lal line. The attenuation due to the CGM+IGM is insufficient to account for the large peak asymmetry (Figure~\ref{fig:spec-B}, right panel), indicating that an outflow structure is present, as is commonly seen in LAEs at $2<z<3$ \citep{Steidel10, Gronke17}.

\subsection{Modelling of the \lal emission line with a shell model}

The outflowing shell model offers a powerful way to extract galaxies' properties from their \lal emission morphology. 
Using the CGM+IGM attenuation curve we just derived, we can now fit the \lal emission profile of \textit{Aerith B} with such an outflowing shell. Although the exact physical meaning of the shell-model is still under debate \citep{Gronke17-meaning,Orlitova18} it is a quick way to extract properties of the scattered medium from \lal spectra. Furthermore, it also accounts for bulk motions affecting e.g.~the asymmetry of the \lal emission line, and is thus more sophisticated than simply measuring the peak separation as done earlier. 

In this simple model, a \lal and continuum-emitting source is surrounded by a dense shell of gas and dust outflowing at constant velocity \citep{Ahn03}. The shell model successfully captures most of the diversity of \lal emission line profiles at both $z\sim0$ \citep{Yang17} and $1<z<3$ \citep{Verhamme08, Verhamme15, Karman17}. It consists of at least 5 free parameters: the bulk velocity $v_{\text{exp}}$ (positive for an outflow), the column density of neutral hydrogen $N_{\text{HI}}$, the gas temperature $T$, the intrinsic width of \lal emission $\sigma_i$, and the optical depth of dust $\tau_d$. 

The modelling and fitting of the shell model is conducted as in \citet{Gronke17} which builds on \citet{Gronke15}. The CGM+IGM attenuation curves discussed in Section 3.3 are applied to reconstruct the \lal profile before absorption. 
In addition to the parameters listed above, we fit the intrinsic equivalent width of the \lal emission line before absorption ($\text{EW}_\text{i}$). The galaxy redshift ($z$) is allowed to vary to optimise the fit; we imposed a Gaussian prior on the systemic redshift $z$ with $(mu,\,\sigma)=(5.793,\,0.003)$ based on the observations which we truncate at $3.5\sigma$. We refer the interested reader to the two papers above for technical details. 

\begin{figure}[t!]\vskip-0.7em
\includegraphics[width=0.5\textwidth]{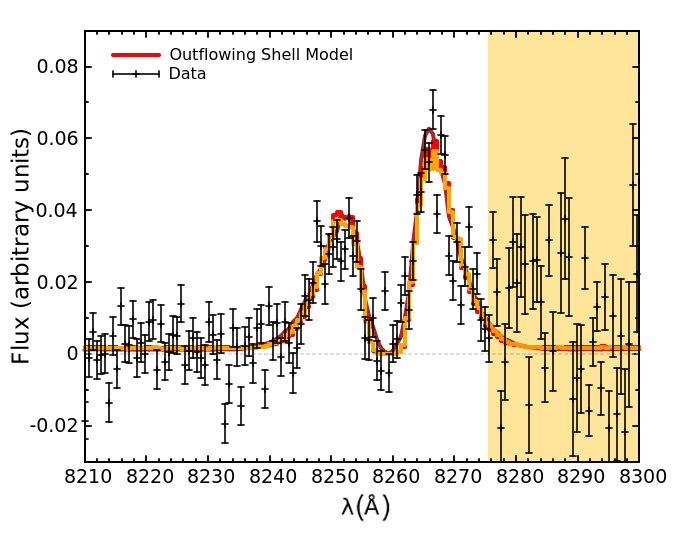}
\caption{Extracted flux of the double-peaked \lal emission line of \textit{Aerith B} (black). The peak separation rules out significant LyC leakage. Red (orange, brown) lines show the best fit outflowing-shell models for three choices of innermost absorption radius $r_\text{min}=20$ ($10, 50$) pkpc (Section 3.4).}
\label{fig:Bfit}
\end{figure}

The curves resulting from the best-fit parameters are shown in Figure~\ref{fig:Bfit}, compared to the observed spectrum. We estimate parameter uncertainties by using the $85\%$ percentiles of the posterior parameter distributions, and include the uncertainty on choice of CGM parameter $r_{\text{min}}$ by running three separate fits for $r_{\text{min}} = 10, 20$ and $50$ pkpc and taking the envelope of the resulting parameter constraints. This yields a hydrogen column density $\text{log} \left(N_{\text{HI}}/\text{cm}^{-2}\right) = 19.3_{-0.2}^{+0.8}$ which is typical of $2<z<3$ LAEs studied similarly in \citet{Gronke17}. The best-fit outflow velocity $v_{\text{exp}} = 16_{-11}^{+4}$ km s$^{-1}$, intrinsic velocity scatter $\sigma_i = 235_{-17}^{+42}$ km s$^{-1}$, and gas temperature $\text{log} T/\text{K} = 3.8_{-0.7}^{+0.8}$, are all typical parameters within $1\sigma$ of those found in lower-$z$ LAEs. We note that this is not driven by the uncertainties on \textit{Aerith B}'s properties, which are $\sim2-3$ times smaller than the intrinsic scatter seen among lower-$z$ objects. The only exception is the dust temperature, $\tau_d = 2.27_{-2.20}^{+2.29}$, whose uncertainty spans the entire range of values observed in $2<z<3$ analogues. We believe this is due to degeneracies with the CGM attenuation curve.

Physical effects beyond our modelling may temper the accuracy of the shell-model fit. For example, the best-fit intrinsic dispersion $\sigma_i$ is a factor $\sim2$ larger than the velocity dispersion of the absorbing gas we assumed in Section~3.3. Ideally, the IGM+CGM attenuation should be modelled at the same time as the outflow. Due to its high column density and extreme external ionization, it is also possible that some of the \lal emission in \textit{Aerith B} comes from \lal fluorescence which is beyond the scope of our modelling (but see discussion in Section 5.3). Our objective was to directly compare the results of the shell-model fitting to the analysis at $2<z<3$ in \citet{Gronke17} by using the same methodology. In conclusion, we found that the physical parameters extracted from \textit{Aerith B} via shell-model fitting are strikingly similar to those seen in $2<z<3$ LAEs in all respects.

\subsection{Correspondence with \lal forest absorption and metallicity}

\begin{figure*}
\centering
\includegraphics[width=0.75\textwidth]{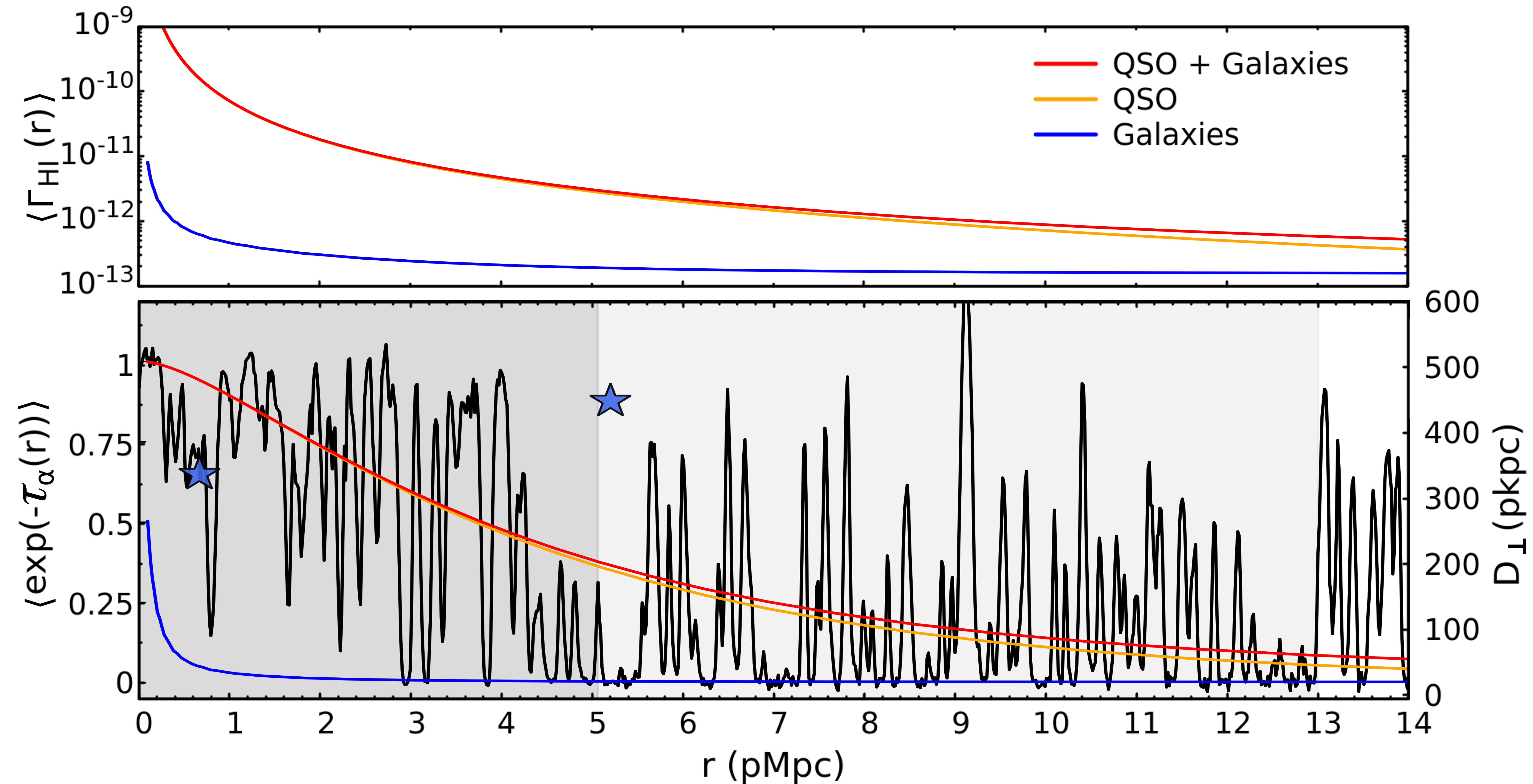}
\vspace{-0.2cm}
\caption{Comparison of the continuum-normalised X-Shooter spectrum of J0836 (black) with the theoretical model of line-of-sight \lal absorption towards the quasar. The model shows the contributions to ({\it top panel}) the photo-ionization rate and to ({\it bottom panel}) the \lal transmission inside the quasar proximity zone, caused by the luminous quasar itself ($M_{\rm UV}=-27.75$, orange) and sub-luminous $M_{\rm UV}>-15$ galaxies clustered around the central quasar (blue). The total effect is indicated in red. The locations of {\it Aerith B} and {\it C} are indicated with stars and the right-hand-side {\it y}-axis gives the angular distance $r_{\perp}$ in proper units. The reported proximity zone sizes by \citet{Eilers17} of $5.06~\rm pMpc$ (dark shaded region) and by \citet{Carilli10} of $13.0\rm~pMpc$ (light shaded region) are also shown.} \label{fig:abs-1D}
\end{figure*}

\textit{Aerith B} appears to coincide in redshift with a \lal absorber inside J0836's proximity zone (Figure~\ref{fig:abs-1D}). Using a high-resolution HIRES \citep{HIRES} spectrum of J0836 first presented in \citet{Bolton11}, we fit this absorber with a Voigt profile using {\tt{vpfit}} \citep{vpfit}. We obtain a column density $\text{log} N_{\text{HI}} = 14.71 \pm 0.05$, too low to constitute a Lyman-limit system (which would require $\text{log} N_{\text{HI}} <17.2$; e.g.~\citealt{Cooper19}). By using the larger wavelength coverage of the X-Shooter spectrum, we search the expected locations of common metal absorbers, finding none. Metallicity limits are obtained by inserting increasingly strong absorbers at those wavelength locations until the absorbing features exceed the spectrum uncertainty, following \citet{Bosman17}. We find abundance [Si/H]$\lesssim -0.01$ and [C/H]$<-0.2$ at $2\sigma$, consistent with low enrichment up to solar. This is consistent with expectations for weak H~{\small{I}} absorbers at $z>3$ (e.g.~\citealt{Fumagalli16}).

The distance between the sightline and \textit{Aerith B} ($D_\perp = 334$ pkpc) is probably too great for this absorber to be associated with the galaxy's CGM (but see \citealt{Rudie12}). However, the feature could be associated with a larger-scale structure such as a gas inflow/outflow, or the CGM of a clustered fainter galaxy.

\section{J0836's proximity zone}

We now demonstrate that a further, independent, valuable aspect of locating proximate LAEs is their utility in constraining the extent and structure of the ionized proximity zones.

The transverse proximity effect detected towards \textit{Aerith A}, the strength of the ionization field at the location of \textit{Aerith B}, and the redshift alignment between \textit{Aerith C} and the end of the proximity zone, all offer constraints on the propagation of ionizing photons from the AGN. Specifically, the observed properties of the proximate LAEs are sensitive to the quasar's opening angle, lifetime or variability, and to the thermal and density profile of the surrounding IGM. 
In this section, we model the impact of these various parameters on J0836's proximity zone together with the \lal emitting galaxies in its environment. 
We closely follow the methodology introduced in \citet{Kakiichi18}, but we extend it to include the quasar radiation field and the visibility of the \lal line in proximate LAEs. 

\subsection{Quasar opening angle}

A quasar shines with an ionizing photon production rate $\dot{N}_{\rm ion}^{\rm QSO}(t)$. 
Using the published broad-band magnitudes of J0836, we measure a spectral slope $\beta=-1.4\pm0.1$ (see Section 5.1) which is consistent with the traditional value of the far-UV spectral energy distribution of quasars, $L_\nu\propto\nu^{-1.5}$ for $\lambda<1050{\rm~\AA}$ \citep{Telfer02}. 
This corresponds to $\dot{N}_{\rm ion}^{\rm QSO}(0)=3.8\times10^{57}\rm~s^{-1}$. We assume that the quasar is radiating in a bipolar cone with an opening angle $\theta_{\rm Q}$. The photo-ionization rate from the quasar in the observed frame is then zero outside the cone, and   
\begin{equation}
    \Gamma_{\rm HI}^{\rm QSO}(r_{\parallel},r_{\perp})=\frac{-\beta \sigma_{912}}{3-\beta}\frac{\dot{N}_{\rm ion}^{\rm QSO}\left[-\Delta t(r_{\parallel},r_{\perp})\right]}{4\pi (r_{\parallel}^2+r_{\perp}^2)}
\end{equation} 
within the cone, where $\sigma_{912}$ is the photo-ionization cross-section at the Lyman limit and $\Delta t$ is the time lag at a distance $(r_{\parallel},r_{\perp})$ from J0836. The distance $r_\parallel$ is the line-of-sight proper distance from the quasar with a positive sign towards the observer and $r_\perp$ is the perpendicular separation along the plane of sky. 

The visibility of a blue \lal peak in {\it Aerith B}, $819\rm~pkpc$ away from J0836, requires a high \lal transparency of the IGM at that distance of at least $\langle e^{-\tau_\alpha(r_{\perp}^B,r_{\parallel}^B)}\rangle\approx80\%$ transmission. 

This necessitates that the galaxy be included within the opening angle of the quasar, which must therefore be larger than
\begin{equation}
\theta_{\rm Q}>\arctan\left(\frac{r_{\perp}^B}{r_{\parallel}^B}\right)\gtrsim24^\circ,
\end{equation}
if the central axis of the bipolar cone is directly pointing toward us, with a strict lower bound $\theta_{\rm Q}>12^\circ$ if we are observing the quasar exactly along the edge of the cone. 
Given J0836 is a very radio-loud quasar \citep{Frey05}, presumably with a jet, we may be observing it closer to the central axis since local observation indicate small intrinsic opening angles of AGN jets with a median of $\theta_{\rm jet} = 1.3^{\circ}$ \citep{Pushkarev17}. Thus we take $\theta_{\rm Q}\geq24^\circ$ as a fiducial constraint.

\subsection{Quasar timescale}

Quasars accreting at or above the Eddington limit, such as J0836 \citep{Kurk07}, often display variability in brightness. The production rate of ionizing photons, and in turn the opacity of the surrounding IGM, will react to AGN variability with a time lag depending on the properties of the IGM and distance from the quasar. Quasar flickering can thus create an ionization ``echo'' in its surroundings. It has long been proposed to use this effect to accurately time the past radiative activity of luminous quasars by using the \lal opacity towards background sources at small impact parameters \citep{Adelberger04, Hennawi06, Visbal08, Schmidt19}. We are now in a position to attempt such a measurement in practice. 
In J0836, we know that the ionizing radiation has reached the location of \textit{Aerith B}, but likely not the location of the slightly more distant \textit{Aerith A} behind the quasar. 

According to the above estimate of the quasar opening angle, \textit{Aerith A} is indeed located within the bipolar cone region unless we are observing J0836 off-axis by more than $>9.7^{\circ}$, which seems disfavoured by its radio-loud nature. Despite this, no continuum is detected immediately bluewards of the \lal emission line in the object's spectrum, despite the fact the redwards continuum is detected at 5.3$\sigma$ (Section 2.3.1). This could imply the ionizing radiation from J0836 has not yet reached \textit{Aerith A} since the onset of the current quasar phase. To determine if this is the case, we must establish (i) the ionization propagation time-lag as a function of distance from the AGN; and (ii) the expected transmission in \textit{Aerith A} if the quasar had been on indefinitely.

\begin{figure}
\includegraphics[width=\columnwidth]{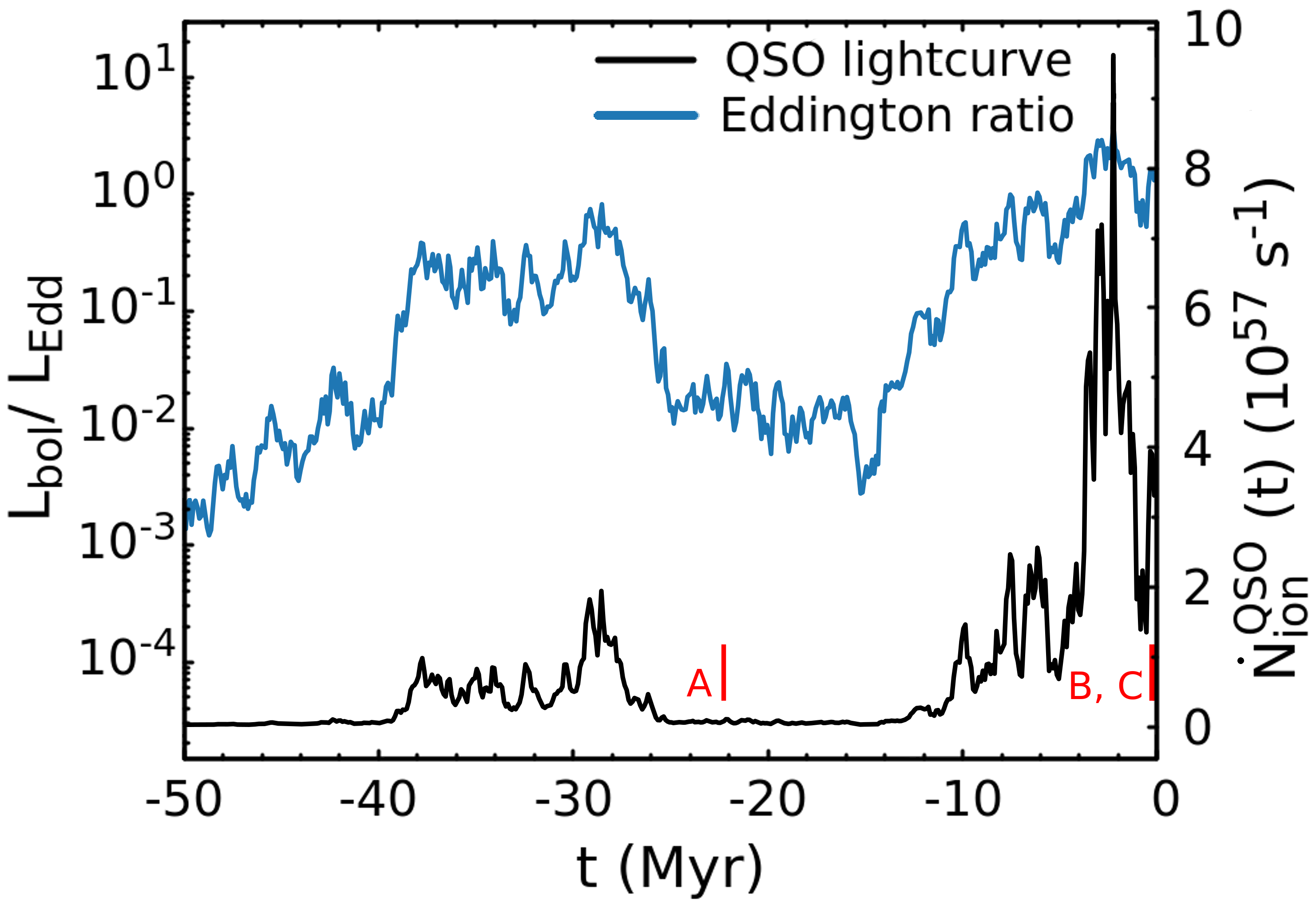}
\caption{A simulated realisation of J0836's lightcurve over the last 50 Myr: (left y-axis: blue) the time-variable Eddington ratio $L_{\rm bol}(t)/L_{\rm Edd}$ and (right y-axis: black) the ionizing photon production rate $\dot{N}_{\rm ion}^{\rm QSO}(t)$ as a function of time. The time $t=0$ corresponds to the time at the quasar redshift $z=5.804$ so that negative values indicate activity at earlier times. The vertical lines (red) mark the time delay surface at the locations of {\it Aerith A}, {\it B}, and {\it C}. }\label{fig:lightcurve}
\end{figure}

\begin{figure}
\hspace*{-0.8cm} 
\includegraphics[width=1.14\columnwidth]{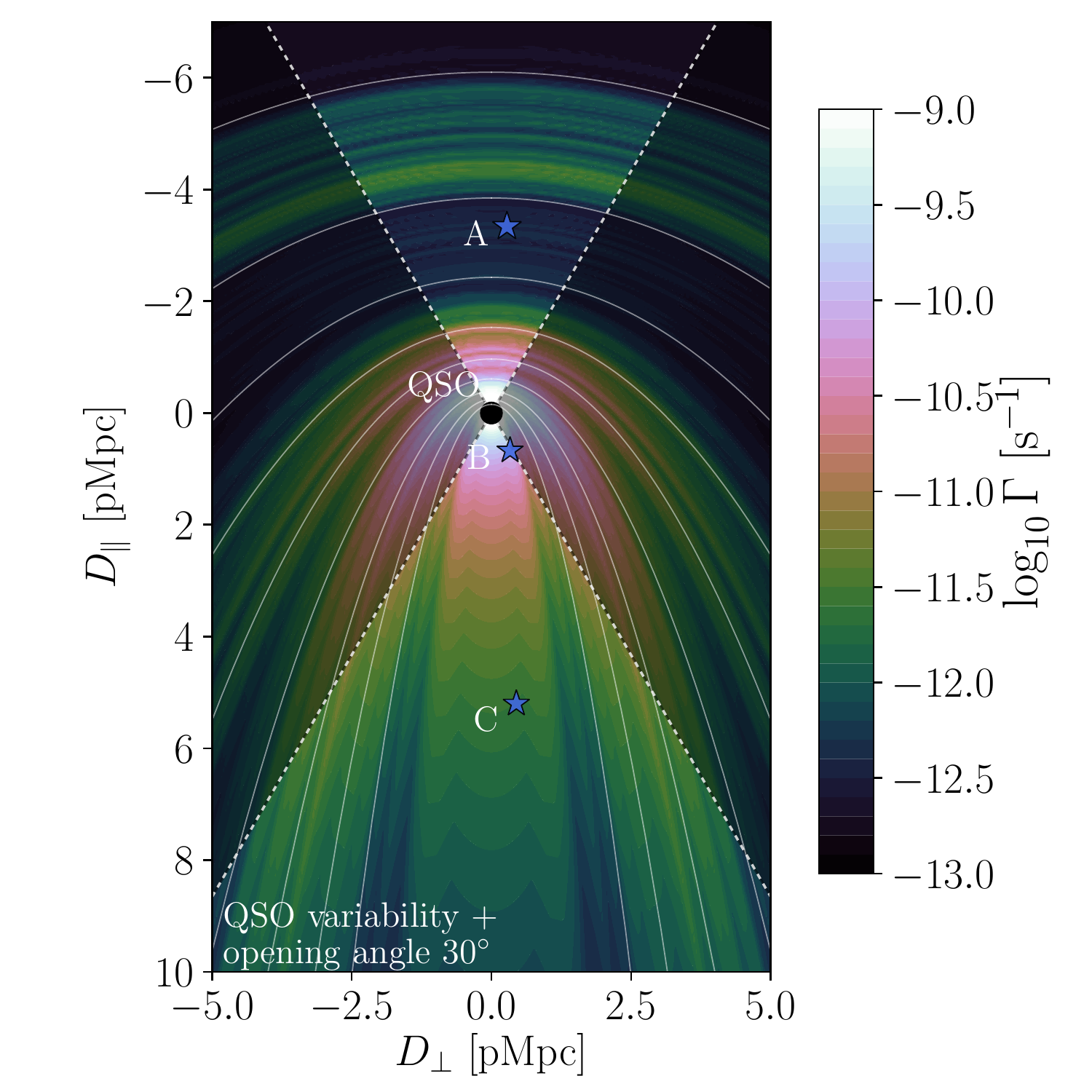}
\caption{Two-dimensional representation of the photo-ionization rate from the quasar in the observed frame. The model corresponds to the lightcurve presented in Figure~\ref{fig:lightcurve} with a biconical opening angle $\theta_{\rm Q}=30^\circ$. The black shaded region indicates the obscured region with $\Gamma_{\rm HI}^{\rm QSO}=0$, but isotropic emission is shown to more easily illustrate the apparent ionizing structure with time delay surfaces. The white contours indicate time delays of $\log(\Delta t/{\rm yr})=6.0,\,6.2,\,6.4,\,6.6,\,6.8,\,7.0,\,7.2,\,7.4,\,7.6$ from inside to outside. The quasar is marked with a circle, and the locations of {\it Aerith A}, {\it B}, and {\it C} relative to the quasar are marked with stars (the relative separations between the LAEs are not correctly captured in this 2D representation). }\label{fig:ion}
\end{figure}

To illustrate the effect of quasar variability, we implement luminosity variations of J0836 into our model using the variable accretion rate shown in blue in Figure~\ref{fig:lightcurve}. It is thought that strong radiatively-driven feedback can halt and regulate the gas fuelling to the central accretion disk or onto the host galaxy (e.g.~\citealt{Hopkins16,Novak11}).
We follow the phenomenological stochastic model of quasar variability by \citet{Kelly09, Kelly14}, which assumes that the quasar light-curve is a realization of damped random walk (also referred to as an Ornstein-Uhlenbeck process; \citealt{Brownian-motion}). We describe the time evolution of the Eddington ratio as
\begin{equation}
d\log\frac{L_{\rm bol}(t)}{L_{\rm Edd}}=\left[\left\langle\log\frac{L_{\rm bol}}{L_{\rm Edd}}\right\rangle-\log\frac{L_{\rm bol}(t)}{L_{\rm Edd}}\right]\frac{dt}{\tau} \\ 
+\sigma_{\rm E} dW(t),
\end{equation}  
where $dW(t)$ is the Gaussian random process and the three parameters, $\left\langle\log L_{\rm bol}/L_{\rm Edd}\right\rangle,~\sigma_{\rm E},~\tau$, correspond to the mean $\log L_{\rm Edd}$, its variability amplitude, and the timescale of variation. For simplicity we assume illustrative values of $\left\langle\log L_{\rm bol}/L_{\rm Edd}\right\rangle=-2$, $\sigma_{\rm E}=0.4$, and a characteristic variability timescale of $\tau=10^7\rm~yr$. 
We take $t=0$ to refer to the time at which J0836 is observed 
leading to $\dot{N}_{\rm ion}^{\rm QSO}(t)\propto L_{\rm bol}(t)/L_{\rm Edd}$ being normalised at $z=5.804$. The response of the quasar luminosity to this varying accretion rate is shown in Figure~\ref{fig:lightcurve}, black. 
The time variability gives rise to a `layered' photo-ionization structure around the quasar as a function of line-of-sight and perpendicular separations indicated in Figure~\ref{fig:ion}, which is directly traceable via the Ly$\alpha$ emission and absorption features of the proximate LAEs and along the line-of-sight absorption of J0836.

We now model the radiation field resulting from this lightcurve following the methodology of \citet{Kakiichi18}. 
The local ionizing background inside the proximity zone includes a contribution from the quasar and from galaxies clustered around the central quasar host (detected and undetected). 
The ionizing power of the three detected galaxies is given by
\begin{equation}
\begin{split}
\Gamma_{\rm HI}\approx4.8\times&10^{-14}\left(\frac{r}{100{\rm~pkpc}}\right)^{-2} \times\\
&\left(\frac{f_{\rm esc}^{\rm LyC}}{0.01}\right) \left(\frac{{\rm SFR}}{20\rm\,M_\odot\,yr^{-1}}\right)\rm\,s^{-1}
\end{split}
\end{equation}
assuming an ionizing emissivity $\xi_{\rm ion}=10^{25.2}\rm~erg^{-1}\,Hz$.
This is a negligible fraction of the collective UV background contribution from the many faint galaxies located more than a few virial radii outside of the host haloes of the LAEs. 
Therefore, we can write the total photo-ionization rate in the quasar's environment as
\begin{equation}
\Gamma_{\rm HI}(r_{\perp},r_\parallel)=\Gamma_{\rm HI}^{\rm QSO}(r_{\perp},r_\parallel)+\langle\Gamma_{\rm HI}^{\rm GAL,CL}(r_{\perp},r_\parallel)\rangle,
\end{equation}
where $\Gamma_{\rm HI}^{\rm QSO}$ and $\langle\Gamma_{\rm HI}^{\rm GAL,CL}\rangle$ are the photo-ionization rates from the quasar and faint (undetected) galaxies surrounding it, respectively. 

We compute the average expected value of the photo-ionization rate due to clustered galaxies using the conditional luminosity function (CLF)-based Halo Occupation Distribution (HOD) framework \citep{Kakiichi18}. 
There have been suggestions of an overdensity of galaxies around J0836 (Z03, A06), indicative of a massive dark matter halo in a biased region. 
We assume a quasar-host halo mass of $M_h>10^{12.5}h^{-1}\rm M_\odot$ and that only the central galaxy is undergoing quasar activity 
The ionizing parameters of the clustered galaxy population are fixed as $\langle f_{\rm esc}\xi_{\rm ion}\rangle=0.10\times10^{25.2}\rm~erg^{-1}Hz$, with a limiting magnitude of contributing galaxies $M_{\rm UV}^{\rm lim}=-15$ \citep{Kakiichi18, Meyer20}.

We are now in a position to calculate the time delay seen by each of our three proximate LAEs. 
If we define $t=0$ to be the time when the light from the quasar reaches to the observer, the observed radiation field at each point in space is sensitive to the quasar luminosity emitted at an earlier time with a time delay $\Delta t$ given by:
\begin{equation}
\Delta t=\frac{(r_{\parallel}^2+r_{\perp}^2)^{1/2}-r_{\parallel}}{c}.
\end{equation}
At the location of the {\it Aerith} galaxies, the visibility of the \lal emission line is sensitive to: 
\begin{enumerate}
\itemsep-0.8em 
\item[] $\Delta t=2.2\times10^7\rm~yr$ for {\it Aerith A}, \vskip0.2em
\item[] $\Delta t=2.3\times10^5\rm~yr$ for {\it Aerith B}, and \vskip0.2em
\item[] $\Delta t=6.2\times10^4\rm~yr$ for {\it Aerith C}
\end{enumerate}
before the time corresponding to J0836's redshift of $z=5.804$ (i.e.~$t = 0.97\rm~Gyr$ after the Big Bang). As {\it Aerith A} is located behind the quasar, by using it as a background source, we can use the \lal forest along its sightline to probe the quasar activity between $t=-2.2\times10^7\rm~yr$ and $t=0$.

\begin{figure}
\hspace*{-0.8cm} 
\includegraphics[width=1.14\columnwidth]{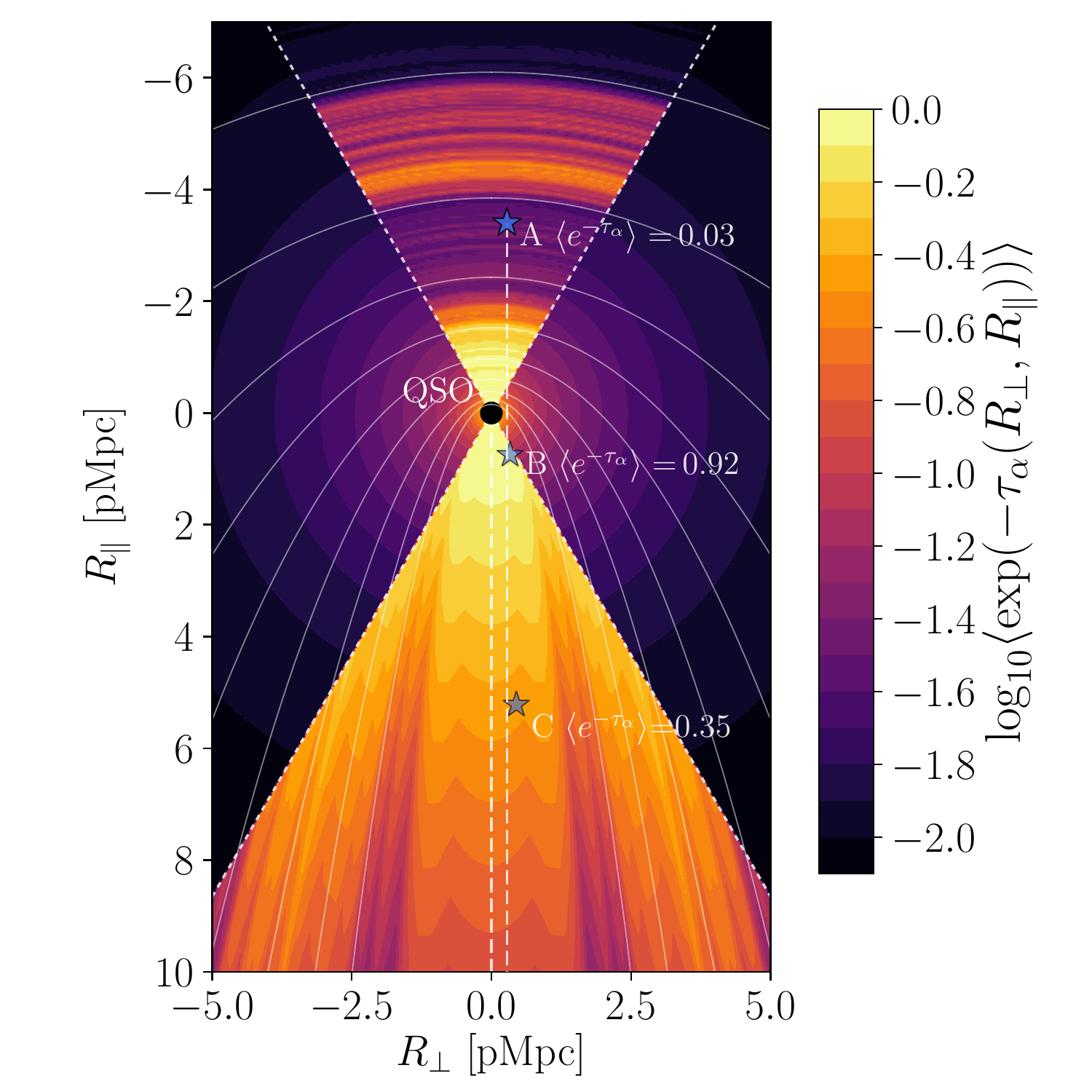}
\vspace{-0.6cm}
\caption{Same as Figure~\ref{fig:ion}, but showing the simulated level of \lal opacity $\langle\exp(-\tau_\alpha)\rangle$ inside J0836's proximity zone as a function of the line-of-sight and perpendicular distances to the quasar. The model includes both the ionizing contributions from the quasar and galaxies surrounding it.}\label{fig:abs-2D}
\end{figure}

Finally, we compute the expected mean \lal absorption along the line-of-sight and transverse directions by convolving the \lal opacity with the probability distribution function of density fluctuations $\Delta_b$,  $P_V(\Delta_b)$ \citep{Pawlik09}; \footnote{We note that the peculiar relative velocity between the IGM gas and the quasar as well as the redshift uncertainties modify the signal along the lines-of-sight, introducing the redshift-space distortions. The redshift evolution further introduces a line-of-sight asymmetry to the 2D \lal absorption map. These are the higher-order effects which we ignore for simplicity.}

\begin{align}
&\langle \exp(-\tau_\alpha(r_{\perp},r_\parallel))\rangle= \nonumber \\
&~~\int d\Delta_b P_V(\Delta_b)\exp\left[-\tau_0\Delta_b^{\beta}\left(\frac{\Gamma_{\rm HI}(r_{\perp},r_\parallel)}{10^{-12}\rm\,s^{-1}}\right)^{-1}\right],\label{eq:model}
\end{align}
where $\beta=2-0.72(\gamma-1)$, 
\begin{equation}
\tau_0\simeq 2.2(1+\chi_{\mbox{\tiny He}})\left(\frac{T_0}{10^4\rm~K}\right)^{-0.72}\left(\frac{1+z}{5}\right)^{9/2} 
\end{equation}
is the optical depth evaluated at mean density (e.g.~\citealt{Becker15-rev}), $\Gamma_{\rm HI} =10^{-12}\rm~s^{-1}$ and $\chi_{\mbox{\tiny He}}$ is the fraction of electrons released by singly ionized ($\chi_{\mbox{\tiny He}}\simeq0.0789$) and doubly ionized ($\chi_{\mbox{\tiny He}}\simeq0.158$) Helium. Based on the measurement from the Doppler widths of \lal absorption lines in J0836, we assume a uniform ($\gamma=1$) temperature at $T_0=1.8\times10^{4}\rm~K$ \citep{Bolton13} and that the helium is doubly ionized. This gives an estimate of the typical proximity zone \lal absorption profile of $z\simeq5.8$ quasars with $M_{\rm UV}=-27.75$ averaged over many realisations of density fluctuations, shown in Figure~\ref{fig:abs-1D}. Clearly, an observed spectrum is modulated around this mean profile because of the fluctuations. In Figure~\ref{fig:abs-2D}, we show the resulting 2D \lal transmission structure inside J0836's proximity zone using the illustrative light-curve described earlier.

In conclusion, we find that if the quasar were active long enough, its radiation field would raise the CGM+IGM transmissivity at the location of \textit{Aerith A} to $T \simeq 55\%$. The observed lack of continuum transmission immediately bluewards of \lal is in mild tension with this prediction ($T<23\%$ over $8305< \lambda <8320$\AA \ at $1\sigma$) but still marginally permitted at $\sim2.4\sigma$. 
In addition, $T = 55\%$ would be sufficient to confidently detect a possible blue peak of the \lal emission line, if it were present and similar to \textit{Aerith B}'s. 
Therefore, J0836's latest quasar phase has likely not lasted long enough to ionize the surroundings of \textit{Aerith A}: the quasar was inactive at least $2.2\times10^7\rm~yr$ ago. This implies the recent active luminous quasar phase has lasted for
\begin{equation}
2.3\times10^5{\rm~yr}<t_{\rm age}<2.2\times10^7\rm~yr,
\end{equation}
where the lower bound is given from the presence of double-peak \lal line in {\it Aerith B}. 

In the past, the quasar duty cycle has been estimated via abundance matching with their dark matter halo masses, which are in turn estimated from the clustering properties of quasars (e.g.~\citealt{Haiman01}). The resulting constraints, $10^6{\rm~yr}<t_{\rm age}<10^9\rm~yr$, are weak yet consistent with our measurement \citep{White12, Conroy13, Cen15}. We are also in agreement with \citet{Eilers17}, who estimate an average quasar episodic lifetime of $t \sim 10^6\rm~yr$ at $z\sim6$ based on the occurrence rate of quasars with very short proximity zones. We discuss the implications of this quasar lifetime on the formation of SMBHs in Section 5.2.


\subsection{Extent and structure of the proximity zone}

In Figure~\ref{fig:abs-1D} we compare the observed continuum-normalized spectrum of J0836 with the \lal absorption model along the sightline of the quasar.
We find that the observed proximity zone size of J0836 \citep{Eilers17} is small for the brightness of the quasar. 
The boundary of the proximity zone is coincident with the location of the foreground LAE \textit{Aerith C}  $\simeq5\rm~pMpc$ in front of the quasar. This suggests the apparent proximity zone size is truncated by Ly$\alpha$ absorption in the CGM of an intervening galaxy. 
If this is the case, the actual size of J0836's proximity zone would be given by the following dip below $10\%$ of the continuum, at $13\rm~pMpc$, as identified by \citet{Carilli10}. This size would be in close agreement with the model expectation, and would be the largest proximity zone ever found around an early quasar - an ionized region stretching $\sim 170\rm~cMpc$ across if it is symmetric on the far side of the quasar.
Some of the other small quasar proximity zone sizes found at $z>5.5$ may be caused by similar Ly$\alpha$ absorption by the intervening CGM of foreground galaxies. 

The contribution of galaxies to the total UV ionization field around the quasar is very sub-dominant to the ionizing radiation from the bright quasar itself (Figure~\ref{fig:abs-1D}) even in the central regions. This is consistent with the view that although quasars turn on in biased environments pre-ionized by galaxy over-densities, once the central galaxy undergoes quasar activity it outshines the surrounding galaxies and dominates the local photo-ionization rate during the quasar lifetime \citep{Lidz06b}. 
Clustered galaxies could still affect the morphology of the proximity zone through higher-order effects not included in our modelling. For example, the increased mean free path of ionizing photons within the biconical ionized region should boost the ionizing power of clustered galaxies in the volume \citep{Davies19-ghost}, which could potentially modulate the transverse extent of the proximity zone perpendicularly to the cone. This motivates the search for associated galaxies in large areas around quasars with extended proximity zones.
 
The amount of transmission observed over the extent of the narrow-band imaging towards \textit{Aerith A} is consistent with the prediction from our model ($\tau_{\text{NB}} = 1.2_{-0.3}^{+0.4}$). This lends credence to the proximity zone extending further than the location of \textit{Aerith C} (Figure\ref{fig:abs-1D}). Furthermore, we expect transmission at the $>90\%$ level at the closest point of intersection of the proximity zone and \textit{Aerith A}'s line-of-sight ($\lambda =8274$\AA), which would be detectable in our DEIMOS spectrum. Sadly, this wavelength is strongly affected by skylines.

\section{Discussion}

\subsection{Lack of quenching}

Early quasars are expected to reside in over-dense regions of the early Universe, both 
due to their rarity and the requirement of continuous inflows of cold gas needed to grow their central SMBH to their observed sizes by $z\sim6$. However, searches for associated galaxy overdensities around early quasars have yielded mixed results (e.g.~\citealt{Kim09, Banados13, Champagne18}) with some quasar fields even appearing to be under-dense \citep{Ota18}. A suggested cause for such under-densities is suppression of star formation by the intense quasar radiation, which can prevent gas from cooling and delay the onset of star formation \citep{Kashikawa07} and even completely photo-evaporate small haloes with $M_h \leq 1 \times 10^7 M_\odot$ on timescales of $\sim 100$ Myr at 1 pMpc distance \citep{Shapiro04}.

Since \textit{Aerith B} is located at $<1$ pMpc from J0836, and the quasar is brighter than those assumed by models in the literature, we wish to determine whether its star formation history could have been significantly impacted. UV intensity is traditionally measured with $J_{21}$, defined as
\begin{equation}
\frac{L(\lambda_{\text{LL}})}{16\pi^2 r^2}  =  J_{21}(r)  \times 10^{-21} \text{erg\ } \text{s}^{-1} \text{cm}^{-2} \text{Hz}^{-1} \text{sr}^{-1},
\end{equation} 
where $L(\lambda_{\text{LL}})$ is the quasar luminosity at the Lyman limit and $r$ is the distance to the quasar. We estimate $L(\lambda_{\text{LL}})$ based on the magnitude of the quasar in the $y_{\text{P1}}$ and $J$ magnitudes measured in \citet{Banados16}, finding a continuum spectral hardness of $\beta=-1.4\pm0.1$ for a Lyman limit luminosity of $L(\lambda_{\text{LL}}) = 3.0 \pm 0.4 \times 10^{32}$ erg s$^{-1}$ Hz$^{-1}$. We obtain UV intensities of $J_{21} = 8.3 \pm 0.9$ at the location of \textit{Aerith C} at the edge of the proximity zone, $J_{21} = 19 \pm 2$ for \textit{Aerith A}, and $J_{21} = 406 \pm 40$ for \textit{Aerith B}.

In the model of \citet{Kashikawa07}, this UV intensity implies that star formation should be suppressed entirely in haloes with $M_h < 3\times10^9 M_\odot$ in J0836's proximity zone. At the location of \textit{Aerith B} only $0.75$ pMpc from the quasar, star formation is suppressed in dark matter haloes with $M_h < 10^{10} M_\odot$ and delayed by $\sim 30$ Myr for haloes of mass $M_h < 3\times 10^{10} M_\odot$. \citet{Chen19} predict a more stochastic suppression of star formation, resulting in a weaker effect overall than \citet{Kashikawa07}, although still qualitatively comparable.

Should star formation in the proximate LAEs have been delayed or even quenched by the quasar's proximity? \citet{Ouchi18} estimated the mass of LAEs at $z\sim5.7$ using a Halo Occupation Distribution (HOD) model, and found that LAEs with $L_{\text{Ly}\alpha} > L_{\text{Ly}\alpha}^* = 6.3 \times 10^{42}$ erg s$^{-1}$ have an average host halo mass of $M = 1.2 \times 10^{11} M_\odot$ and a minimum mass of $M_{\text{min}} = 3.5 \times 10^9 M_\odot$. \textit{Aerith B} and \textit{C} are brighter than $L_{\text{Ly}\alpha}^*$ even discounting the extra emission in the blue peak of the \lal line, implying their star formation history would not be impacted if they reside in average-sized dark matter haloes. However, the lack of quenching implies a host halo mass of $M_h > 3\times 10^{10} M_\odot$ for \textit{Aerith B}, more stringent than the lower limit from the HOD model. \textit{Aerith A} has a luminosity below $L_{\text{Ly}\alpha}^*$ (Table~\ref{table:megatable}), but lack of quenching still implies a host halo mass larger than $\sim10^{10} M_\odot$. 

On average, one would expect a strong suppression of galaxies with $L_{\text{Ly}\alpha}<L_{\text{Ly}\alpha}^*$ inside J0836's proximity zone. This is tempered by variability both of the LAEs and the quasar. The degree of suppression of star formation in proximate galaxies depends on the status of star-forming processes at the time of quasar turn-on (cold gas reservoirs, ISM properties) as well as the duration of the current quasar phase and cosmic variance \citep{Habouzit19, Chen19}. Nevertheless, the field around J0836 is the ideal laboratory for testing quasar radiative suppression, as it will affect bright LAEs within current observational reach over a very large surface area of sky. Future analysis and observations will confirm whether the proximity zone is indeed over-dense in bright galaxies, and under-dense in fainter ones, as predicted by models.

\subsection{Implications for the formation of $z\gtrsim6$ SMBHs: accretion disks \& the host-galaxy environment}

The growth of SMBHs with masses $\sim10^9\rm~M_\odot$ by $z\sim6$ requires substantial gas accretion or massive seeds such as those produced by the direct collapse scenario or the collapse of dense star clusters (see e.g.~\citealt{Haiman13,Woods19,Smith19} for reviews). The $e$-folding accretion timescale for the black hole mass growth is (e.g.~\citealt{Madau14}) 
\begin{equation}
t_{\rm acc}^{\rm BH}=3.8\times10^8\frac{\epsilon}{1-\epsilon}\left(\frac{L_{\rm bol}}{L_{\rm  Edd}}\right)^{-1}\rm~yr.
\end{equation}
For the conventional value of radiative efficiency, $\epsilon=0.10$, assuming a thin accretion disc \citep{Shakura73} the timescale is $t_{\rm acc}^{\rm BH}=4.2\times10^7\rm~yr$ at the Eddington limit. 

An alternative estimate of the total quasar lifetime follows from the quasar clustering and luminosity function measurements, which constrain the population-averaged duty cycle $f_{\rm duty}$ (e.g.~\citealt{White12}, see Section 4.2). The average total lifetime of quasar activity at $z=5.8$ is then 
\begin{equation}
\langle t_{\rm Q}\rangle\approx\frac{f_{\rm duty}t_{\rm H}(z=z_{\rm Q})}{1-\cos\theta_{\rm Q}}\leq1.12\times10^8{\rm yr}\left(\frac{f_{\rm duty}}{0.1}\right),
\end{equation}
using the bound of opening angle measured from the J0836 field. $f_{\rm duty}$ is uncertain by an order of magnitude but 0.1 is somewhat upper estimate. 

Our estimated age of the recent quasar activity of J0836 is clearly shorter than the $e$-folding timescale required to grow to a SMBH of $M_{\rm BH}\approx2\times10^9\rm~M_\odot$ and much shorter that the total quasar lifetime. This implies transient, episodic super-Eddington gas accretion but with modest quasar activities to match the observed state of J0836, or sustained accretion in obscured phases.

Such fast intermittent gas accretion with quasar episodes of mildly super-Eddington luminosities could be a consequence of {\it slim} accretion discs around high-redshift SMBHs (see \citealt{Madau14}) where the radiation-dominated, advective flow of the disk naturally lowers the radiative efficiency ($\epsilon\sim0.02$ \& $L/L_{\rm Edd}\sim1$) for super-Eddington accretion ($\dot{M}/\dot{M}_{\rm BH}>1$), resulting in only mildly super-Eddington luminosity. The recent (general-relativistic) radiation magneto-hydrodynamic simulations of super-Eddington accretion disks \citep{McKinney14,Jiang14,Sadowski16} indicate that such scenario arises as long as a vigorous inflow and a reservoir of gas around the disk is maintained.

The Mpc-scale overdensity around J0836 indicated by the bright proximate LAEs supports that such a reservoir of gas around the quasar-host galaxy could be maintained, potentially triggering the occasional nuclear inflow by mergers or sustained cold accretion onto the galaxy. \citet{Frey10} suggest that the compact observed radio size ($\sim\,$$40\rm\,pc$) of J0836 and its steep spectral index could result from interactions of a relativistic jet with the dense environment of the host galaxy, giving rise to emission peaked at frequencies of a few GHz \citep{Falcke04}.

For comparison, $z\sim3$ studies of transverse and line-of-sight He~{\small{II}} proximity effects \citep{Schmidt17} and \lal fluorescence sources around hyper-luminous quasars \citep{Hennawi13, Trainor13,Borisova16} also indicate short durations of the radiative activity of quasars on $\sim10^{6-7}\rm~yr$ timescales. \citet{Eilers17} also find small sizes of \lal proximity zones in $z\sim6$ quasars. While many uncertainties remain, such findings seem to align with the above picture. Spectroscopic follow-up of quasar fields, piloted with J0836, shows a clear way forward demonstrating the capability of proximate LAEs as a laboratory to constrain the formation and growth of the first SMBHs.

\subsection{Possible effect of \lal fluorescence}

The \lal luminosities of the proximate LAEs could include a contribution from \lal fluorescence caused by the quasar's ionization field, in addition to star formation (e.g.~\citealt{Cantalupo05, Hennawi13}). 
The amount of \lal fluorescence imparted on a system with $\HI$ column density $N_{\rm HI}$ and cross section $\sigma_{\rm LAE}$ at a distance $r$ away from the quasar is given by
\begin{align}
L_{\alpha}^{\rm fl.}&=\frac{2}{3}h\nu_\alpha f_{\phi}\int_{\nu_{\rm L}}^\infty(1-e^{-\sigma_\nu\NHI})\frac{\sigma_{\rm LAE}}{4\pi r^2}\frac{L_{\nu}^{\rm QSO}}{h\nu}d\nu, \\
&\simeq\frac{2}{3}h\nu_\alpha f_\phi\frac{\sigma_{\rm LAE}}{4\pi r^2}\dot{N}_{\rm ion}^{\rm QSO},~\mbox{for $\NHI\gg10^{17}\rm~cm^{-2}$}, \nonumber
\end{align}
where $f_{\phi}$ is the illumination fraction.\footnote{For a spherical cloud, the illumination fraction is $f_{\phi}=1-\phi/\pi$ where $\phi$ is the phase angle between the quasar-cloud-observer, having $f_{\phi}=0$, $\phi=\pi$ if all fluorescent \lal is backscattered away from us, $f_{\phi}=0.5$ for half-moon ($\phi=\pi/2$) illumination, and $\phi=0$ if the fluorescent \lal is scattered into out line of sight. For a cloud consisting of tiny cloudlet structures, the fluorescent \lal can be scattered within the cloud and redirected to us. Thus, the value of $f_{\phi}$ can become larger than a solid spherical geometry. Here we assume the conservative upper bound contribution of $f_{\phi}=1$.} Assuming the geometrical cross section $\sigma_{\rm LAE}=\pi r_{\rm size}^2$ with radius $r_{\rm size}$, the fluorescent \lal luminosity in our objects is
\begin{equation}
L_{\alpha}^{\rm fl.}\approx2.7\times10^{41}r_{\rm Mpc}^{-2}\left(\frac{r_{\rm size}}{5\rm~kpc}\right)^2\left(\frac{\dot{N}_{\rm ion}^{\rm QSO}}{4\times10^{57}\rm s^{-1}}\right)\rm erg~s^{-1}.
\end{equation}
The size of a LAE is typically $\lesssim2\rm~pkpc$ \citep{Shibuya19}; we adopt a value of $5 \text{pkpc}$ as an upper estimate.

The above calculation indicates that \lal fluorescence accounts for at most $5\%$ of the observed luminosity of \textit{Aerith B}, and $<1\%$ for \textit{Aerith A} and \textit{C}. The \lal luminosities of the objects are in close agreement with the expected values based on their UV luminosities and star-formation rates. Therefore, the fluorescent \lal contribution due to J0836's ionizing field is minor.

\subsection{Future prospects}

While the serendipitous discovery of our proximate LAEs was unquestionably aided by the extreme luminosity of J0836, there is no reason why such couldn't be found at smaller separations from less bright quasars. Over 30 (and growing) fields around $z>5.7$ quasars have been observed with integral field spectroscopy by the MUSE instrument \citep{Farina19}. Finding additional proximate LAEs offers a clear path to refining the results presented in this paper.

With a larger sample, proximate LAEs will constrain models of reionisation. Faint galaxies are expected to provide the majority of reionisation photons, but scenarios driven by bright galaxies are still permitted \citep{Meyer20}. In the model of \citet{Naidu19}, galaxies brighter than $M_{\text{UV}} = -20$ can power reionization alone if their escapes fractions are $f_{\rm esc} \geq 0.20$. Proximate LAEs provide the most direct way of measuring $f_{\rm esc}$ in individual bright and faint galaxies, and will eventually rule in favour or against existing predictions. In addition, numerical models are resolving the physical properties of galaxies in the first billion years in increasingly fine detail (e.g.~\citealt{Pallottini17}). It will be interesting to see if those frameworks can account for LyC leakage from some bright galaxies during the EoR, but not others.

Some model of the neutral IGM's effect on LAEs predict that the red peak of Ly-$\alpha$, in addition to the blue peak, will be suppressed on average by a damping wing of neutral hydrogen absorption (e.g.~\citealt{Dijkstra07}). It is interesting to note that our 3 galaxies, although presumably less affected by this than the general $z\sim5.8$ LAE population, are slightly faint in \lal for their $M_{\text{UV}}$ (Figure~\ref{fig:maglum}). In the future, larger samples of proximate LAEs may make it possible to reconstruct an `intrinsic' \lal luminosity distribution function and disentangle changes in the IGM and the CMG of early galaxies. At $z>6$, this may offer a path to measuring the IGM damping wing in a statistical sense.

\section{Summary} 

We have discovered the first three proximate LAEs in the proximity zone of quasar \quasar at $z=5.8$. The intense ionizing radiation in their surroundings reveals unique properties never observed before at $z>5$. To understand these observations in the context of the central quasar, we have modelled the ionization and density structure of the proximity zone and put constraints on the central quasar's properties. Our main findings are as follows:
\begin{itemize}
	\item \textit{Aerith B} displays the first widely-separated \lal emission line seen in a galaxy during reionization. Unlike previously detected \lal double peaks at $z>6.5$, the morphology and luminosity of \textit{Aerith B}'s \lal line is normal and even typical compared to $2<z<3$ LAEs.
	\item The morphology of the \lal line in \textit{Aerith B} implies an escape fraction of ionizing radiation $f_{\text{esc}} \simeq 0.01$ based on the double peak separation/LyC leakage correlation, which is well-calibrated at low $z$. This implies that not all bright galaxies ($M_{\text{UV}} = -21$) during the EoR are strong leakers, in tension with some models of reionization.
	\item Fitting the \lal line morphology with an outflowing shell model, we find best-fit ISM properties of \textit{Aerith B} implying a typical central $N_{\text{H}}$ density, gas outflow speed, and dust opacity compared to the $2<z<3$ LAE population.
	\item The star-formation in \textit{Aerith B} has not been quenched, despite being exposed to an ionizing intensity of $J_{21} = 406$. This might imply it is hosted in a dark matter halo with mass $M_h > 3 \times 10^{10} M_\odot$, or that it assembled its stellar mass before quasar turn-on.
	\item \textit{Aerith A} is detected at $10\sigma$ in a narrow-band filter at a wavelength shorter than Ly-$\alpha$. The covered wavelength range significantly overlaps with J0836's proximity zone, making this the first detection of the \lal transverse proximity effect. Modelling the propagation of ionizing photons around J0836, we find the level of transmission towards \textit{Aerith A} ($\tau_{\text{NB}} = 1.2_{-0.3}^{+0.4}$) to be entirely consistent with the quasar's UV magnitude assuming a far-UV slope of $\beta=-1.4$. The ionizing emission of J0836 is therefore not significantly obscured.
	\item J0836's current quasar phase started $2.3\times10^5{\rm~yr}<t_{\rm age}<2.2\times10^7\rm~yr$ ago, based on the lack of continuum transmission immediately bluewards of \textit{Aerith A}'s \lal line ($2.4\sigma$). This is the first measurement of its kind, and is consistent with theoretical expectations from the literature.
	\item \textit{Aerith C} displays an emission line which lines up closely ($\Delta z = 0.02$) with the formal end of J0836's proximity zone. The following distance at which transmission falls below $10\%$ is $r=13$ pMpc, in close agreement expectations from our model based on J0836's UV magnitude. Such an extended proximity zone would be the largest ever detected around an EoR quasar.
	\item The UV continuum in \textit{Aerith C} appears to be offset from the location of the detected emission line by $6.3$pkpc. Such an offset is larger than previously observed at high $z$, suggesting a larger overdensity of galaxies.
\end{itemize}

\section*{Acknowledgements}

SB, KK, RM and NL acknowledge funding from the European Research Council (ERC) under the European Union's Horizon 2020 research and innovation programme (grant agreement No.~669253).
MG was supported by by NASA through the NASA Hubble Fellowship grant HST-HF2-51409. 
Based in part on data collected at Subaru Telescope, which is operated by the National Astronomical Observatory of Japan.
Some of the data presented herein were obtained at the W.~M.~Keck Observatory, which is operated as a scientific partnership among the California Institute of Technology, the University of California and the National Aeronautics and Space Administration. The Observatory was made possible by the generous financial support of the W.~M.~Keck Foundation. 

The authors recognize and acknowledge the very significant cultural role and reverence that the summit of Maunakea has always had within the indigenous Hawaiian community.  We are most fortunate to have the opportunity to conduct observations from this mountain.

\facility{Keck:II (DEIMOS)}

\bibliography{bibliography}



\end{document}